\documentclass{aa}
\usepackage{epsfig,times}
\usepackage{color}

\begin{document}
\thesaurus{12     % A&A Sect 12: Physical and chemical processes
      (02.13.2;   % MHD
       02.01.2;   % Accretion, accretion disks
       09.10.1;   % ISM jets and outflows
       08.13.1;   % Stars: magnetic fields
       08.13.2;   % Stars: mass loss
       08.16.5) } % Stars: pre-main sequence
\title{Long-term evolution of a dipole type magnetosphere interacting
with an accretion disk.}
\subtitle{II. Transition into a quasi-stationary spherically radial
outflow}
 
\author{Christian Fendt \and Detlef Elstner}
 
\institute{
Astrophysikalisches Institut Potsdam, An der Sternwarte 16,
D-14482 Potsdam, Germany; email: cfendt@aip.de ; delstner@aip.de
}

\date{Received ??; accepted ??}

\authorrunning{Fendt \& Elstner}
\titlerunning{Evolution of dipole type magnetospheres with disks} 

\maketitle

\def\ri{r_{\rm i}}
\def\ro{r_{\rm out}}
\def\rs{r_{\star}}
\def\mj{\dot{M}_{\rm jet}}
\def\ms{\dot{M}_{\star}}
\def\vk{v_{\rm K}}
\def\ti{t_{\rm i}}

\def\rj{R_{\rm jet}}
\def\rk{R_{\kappa}}

\def\bp{B_{\rm P}}
\def\bh{B_{\phi}}
\def\jp{\vec{B}_{\rm P}}

\def\cm{{\rm cm}}
\def\msun{{\rm M}_{\sun}}
\def\rsun{{\rm R}_{\sun}}

\begin{abstract}

The evolution of an initially stellar dipole type magnetosphere 
interacting with an accretion disk is investigated numerically using 
the ideal MHD ZEUS-3D code in the 2D-axisymmetry option.
Depending mainly on the magnetic field strength, our simulations
may last several thousands of Keplerian periods of the inner disk.
A Keplerian disk is assumed as a boundary condition prescribing a
mass inflow into the corona.
Additionally, a stellar wind from a rotating central star is
prescribed.
We compute the innermost region around the stellar object applying
a non-smoothed gravitational potential.

Our major result is that the initially dipole type field develops
into a spherically radial outflow pattern with two main components, a
disk wind and a stellar wind component.
These components evolve into a quasi-stationary final state.
The poloidal field lines follow a conical distribution.
As a consequence of the initial dipole,
the field direction in the stellar wind is opposite
to that in the disk wind. 
The half opening angle of the stellar wind cone varies from $30\degr$ to
$55\degr$ depending on the ratio of the mass flow rates of disk
wind and stellar wind.
The maximum speed of the outflow is about the Keplerian speed at the
inner disk radius. 

An expanding bubble of hot, low density gas together with the 
winding-up process due to differential rotation between star
and disk disrupts the initial dipole type field structure. 
An axial jet forms during the first tens of disk/star rotations,
however, this feature does not survive on the very
long time scale.
A neutral field line divides the stellar wind from the disk wind.
Depending on the numerical resolution, small plasmoids are 
ejected in irregular time intervals along this field line.
Within a cone of $15\degr$ along the axis the formation of 
small knots can be observed if only a weak stellar wind is
present.

With the chosen mass flow rates and field strength we see
almost no indication for a flow self-collimation. 
This is due to the small net poloidal electric current in the 
(reversed field) magnetosphere which is in difference to typical
jet models.

%---------------------------------------------------------------

	      
\end{abstract}

%---------------------------------------------------------------
\section{Introduction}
%---------------------------------------------------------------
%
A stellar dipole type magnetic field surrounded by an accretion
disk is the common model scenario for a variety of astrophysical
objects.
Examples are the classical T\,Tauri stars, 
magnetic white dwarfs (in cataclysmic variables)
and neutron stars (in high mass X-ray binaries).
Part of these sources exhibit Doppler shifted emission lines 
indicating wind motion.
Highly collimated jets have been observed from young stellar objects 
and X-ray binaries, where also quasi-periodic oscillations (QPO) are 
observed.
In general, magnetic fields are thought to play the leading role
for the jet acceleration and collimation 
(Blandford \& Payne 1982; Pudritz \& Norman 1983; 
Camenzind 1990; Shu et al. 1994; Fendt et al. 1995).

Recently, several papers were considering the evolution of a stellar
magnetic dipole in interaction with a diffusive accretion disk.
Hayashi et al. (1996) observed magnetic reconnection and the
evolution of X-ray flares during the first rotational periods.
Miller \& Stone (1997) Goodson et al. (1997) included the evolution
of the (diffusive) disk structure in their calculation.
In these papers a collapse of the inner disk is indicated depending on
the magnetic field strength and distribution.
The inward accretion flow develops a shock near the star.
The stream becomes deflected resulting in a high-speed flow in axial
direction.

The results of Goodson et al. (1997, 1999) and Goodson \& Winglee (1999)
are especially interesting since combining a huge spatial scale
(2\,AU) with a high spatial resolution near the star ($0.1\rsun$).
However, to our understanding it is not clear, how the initial condition
(a standard $\alpha$-viscosity disk) is really developing in their code
without any physical viscosity.
Secondly, not very much is said about the amount of magnetic diffusivity.
The assumption of constant diffusivity cannot really reflect the two
component model of disk and coronal out flow.

Time-dependent simulations lasting only a short time scale 
depend strongly on the initial condition
and the calculation of the evolution of such a magnetosphere over
{\em many} rotational periods is an essential step.
In particular, this is important point if the initial condition is not
in equilibrium.
In summary, we note that all calculations including the treatment of the
disk structure could be performed only for a few Keplerian periods of the
inner disk (and a fraction of that for the outer disk).

At this point, we emphasize that the observed kinematic time scale of 
protostellar jets can be as large as $10^3 - 10^4$\,yrs,
corresponding to $5\times 10^4 - 5\times 10^5$ stellar rotational periods 
(and inner disk rotations)!
For example, proper motion measurements for the HH30 jet 
(Burrows et al. 1996)
give a knot velocity of about $100-300\,{\rm km\,s^{-1}}$ and a 
knot production rate of about 0.4 knot per year.
Assuming a similar jet velocity along the whole jet extending
along 0.25\,pc (Lopez et al. 1995),
the kinematic age is about 1000 yrs.

A different approach for the simulation of magnetized winds from accretion 
disks 
considers the accretion disk `only' as a {\em boundary condition} for the
mass inflow into the corona.
Since the disk structure itself is not treated, such simulations may last
over hundreds of Keplerian periods.
This idea was first applied by Ustyugova et al. (1995).
Extending this work, Romanova et al. (1997) found a stationary final state
of a slowly collimating disk wind in the case of a split-monopole initial
field structure after 100 Keplerian periods.
Ouyed \& Pudritz (1997, hereafter OP97) presented time-dependent simulations 
of the jet formation from a Keplerian disk. 
For a certain (already collimating) initial magnetic field distribution,
a stationary state of the jet flow was obtained after about 400
Keplerian periods of the inner disk with an increased degree of
collimation.
In a recent extension of their work both groups were considering the 
influence of the grid's shape on the degree of collimation 
(Ustyugova et al. 1999)
and the effect of the mass flow rate (Ouyed \& Pudritz 1999).
Ouyed \& Pudritz (2000) investigate the problem of jet stability and 
magnetic collimation extending the axisymmetric simulations to
3D.

In this paper, we are essentially interested in the evolution of the
ideal MHD magnetosphere and the formation of winds and jets
and not in the evolution of the disk structure itself.
Therefore, we do not include magnetic diffusivity into our simulations.
The disk acts only as a boundary condition for the corona/jet region.
The winding-up process of poloidal magnetic field due to strong
differential rotation between the star and the disk would always be
present even if a disk diffusivity would have been taken into
account. 
The disk diffusivity will never be so large that a rigid rotation of
the magnetosphere in connection with the disk can be maintained.

In that sense our simulations represent an extension of the OP97
model, taking additionally into account the central star as a
boundary condition and a stellar dipole type field as initial
condition. 
First results of our simulations were presented in 
Fendt \& Elstner (1999, hereafter FE99).
Here we give a more detailed discussion together with new results.
Movies of our simulations will be provided under 
{\sl http://www.aip.de/$\sim$cfendt}.

%---------------------------------------------------------------
\section{Basic equations}
%---------------------------------------------------------------
%
Using the ZEUS-3D MHD code 
(Stone \& Norman 1992a,b; Hawley \& Stone 1995)
in the 2D-axisymmetry option we solve the system of 
time-dependent ideal magnetohydrodynamic equations,
\begin{equation}
\frac{\partial \rho}{\partial t} + \nabla \cdot (\rho \vec{v} ) = 0\,,
\end{equation} 
\begin{equation}
\frac{\partial\vec{B} }{\partial t} -
\nabla \times (\vec{v} \times \vec{B}) = 0\,,
\end{equation} 
\begin{equation}
\nabla \cdot\vec{B} = 0\,,
\end{equation} 
\begin{equation}
\rho \left[ \frac{\partial\vec{v} }{\partial t} 
+ \left(\vec{v} \cdot \nabla\right)\vec{v} \right]
+ \nabla (P + P_{\rm A}) + \rho\nabla\Phi - \vec{j} \times \vec{B} = 0\,,
\end{equation} 
%
%
%\begin{equation}
%\rho \left[ \frac{e}{\partial t} 
%+ \left(\vec{v} \cdot \nabla\right)e \right]
%+ P (\nabla \cdot\vec{v} ) = 0\,,
%\end{equation} 
%
where $\vec{B}$ is the magnetic field,
$\vec{v}$ the velocity,
$\rho $ the gas density,
$P $ the gas pressure,
%$e $ the internal energy,
$\vec{j} = \nabla \times \vec{B} / 4\pi$ the electric current density.
and $\Phi$ the gravitational potential.
We assume a polytropic ideal gas, $P = K \rho^{\gamma}$ with 
a polytropic index $\gamma = 5/3$.
%Therefore we do not solve the energy equation (4).
Similar to OP97, we have introduced a turbulent magnetic pressure 
due to Alfv\'en waves, $P_{\rm A}\equiv P/\beta_T $,
where $\beta_T $ is assumed to be constant.
OP97 considered the turbulent magnetic pressure in order to 
support the cold corona of e.g. young stellar accretion disks
for a given gas pressure.
Clearly, the assumption of a {\em constant} $\beta_T $ is motivated
by the reason of simplification.
Using dimensionless variables,
$r' \equiv r/r_i $,
$z' \equiv z/z_i $,
$v' \equiv v/v_{K,i} $,
$t' \equiv t r_i / v_{K,i}$,
$\rho' \equiv \rho/\rho_i $,
$P' \equiv P/P_i $,
$B' \equiv B/B_i $,
$\Phi' = - 1/\sqrt{r'^2 + z'^2} $',
where the index $i$ refers to parameter values at the inner disk 
radius $r_i$, 
the normalized equation of motion eventually being solved with the
code is 
\begin{equation}
\frac{\partial\vec{v}' }{\partial t'} 
+ \left(\vec{v}' \cdot \nabla'\right)\vec{v}' = 
\frac{2 \,\vec{j}' \times \vec{B}' }{\delta_i\,\beta_i\,\rho'}
- \frac{\nabla' (P' + P_{\rm A}')}{\delta_i \,\rho'} - \nabla'\Phi'\,.
\end{equation} 
The coefficients 
$\beta_i \equiv 8 \pi P_i / B_i^2 $ and
$\delta_i \equiv \rho_i v^2_{K,i} / P_i $ with 
the Keplerian speed $v_{K,i} \equiv \sqrt{GM/r_i} $,
correspond to the plasma beta and the Mach number of the 
rotating gas.
For a `cold' corona with $P_{\rm A}' > 0$, it follows 
$\beta_T = 1 / (\delta_i (\gamma - 1) / \gamma - 1)$.
In the following we will omit the primes and will discuss only
normalized variables if not explicitly declared otherwise.

Note that in our figures the horizontal axis is always the $z$-axis
and the vertical axis is the $r$-axis.

%---------------------------------------------------------------
\section{The model -- numerical realization}
%---------------------------------------------------------------
%
In general, our model represents a system consisting
of a central star and an accretion disk separated by a gap.
Star and disk are initially connected by an dipole type
magnetosphere. Axisymmetry is assumed.
The stellar rotational period can be chosen arbitrarily.
The disk is in Keplerian rotation.
Disk and star are taken into account as an inflow boundary
condition. 
It has the advantage that the behavior of the wind flow can be
studied independently from the evolution of the accretion disk.

This is an essential point, since the numerical simulation of 
the magnetized disk structure represents itself one of the most 
complicated and yet unresolved problems of astrophysics.
It is therefore unlikely to find a proper disk initial condition
which is in equilibrium.
Yet, all MHD disk simulations could be performed only for a few
Keplerian periods 
(e.g. Hayashi et al. 1996, Miller \& Stone 1997; Kudoh et al. 1998).
A global solution of the disk-jet evolution does not yet seem to be
numerically feasible.

The general disadvantage involved with such a fixed disk (plus star) 
boundary condition is 
that the fundamental question of the wind/jet formation evolving
out of the accretion disk cannot be investigated.

%---------------------------------------------------------------
\subsection{Numerical grid and initial condition}
Prescribing a stable force-equilibrium as initial condition is 
essential for any numerical simulation.
Otherwise the simulation will just reflect the relaxation process
of such an unstable (and therefore arbitrary) initial condition
to a state of stability.
In particular this would be important if only few time steps are
computed.

In our model, 
we assume an initially force-free (and also current-free) magnetic 
field together with a density stratification in hydrostatic 
equilibrium.
Such a configuration will remain in its initial state if not
disturbed by a boundary condition.
The initial density distribution is $\rho(r,z) = (r^2 + z^2)^{-3/4}$.
The gravitational point mass is located half a grid element below the
origin.
Due to our choice of cylindrical coordinates we cannot treat the
stellar surface as a sphere. 
Along the {\em straight} lower $r$-boundary we define the 
'stellar surface' from $r=0$ to $r=\rs$, 
a gap from $r=\rs$ to $r=r_i = 1.0$,
and the disk from  $r=1.0$ to $r=r_{\rm out}$ (see Fig.\,1).

We have chosen an initial field distribution of a force-free, current-free
magnetic dipole,
artificially deformed by 
(i) the effect of 'dragging' of an accretion disk, and
(ii) an 'opening' of the field lines close to the outflow boundaries.
The initial field distribution is calculated using a stationary finite
element code described in Fendt et al. (1995).
In this approach, the axisymmetric $\phi$-component of the vector potential
$A_{\phi}$ is computed 
(as solution of the well-known Grad-Shafranov equation)
using a numerical grid with twice the resolution of the grid applied
in the ZEUS code.
Our finite element code allows for a solution of the stationary
boundary value problem for {\em any} boundary condition.
Thus, we are able to define any force-free solution as initial condition
for the simulation.

With that, from the vector potential the initial field distribution for
the time-dependent simulation is derived with respect to the ZEUS-3D
staggered mesh,
\begin{eqnarray}
  B_z(i,j) & = & \ 2\,(A_{\phi; i,j+1} - A_{\phi; i,j}) /
  (r^2_{i,j+1} - r^2_{i,j})\,, \nonumber  \\
  B_r(i,j) & = &  - \left(A_{\phi; i+1,j} - A_{\phi; i,j}\right)/
   \left(r_{i,j}\,(r_{i+1,j} - r_{i,j})\right).
\end{eqnarray}
Here, the first and second index $i$ and $j$ denote the 
$z$- and $r$-direction, respectively.
A suitable normalization factor is multiplied in order to match the
field strength defined by the coefficients $\beta_i$ and $\delta_i$.
With this approach the maximum normalized $|\nabla \cdot\vec{B}|$ 
is $10^{-15}$ and $|\vec{j} \times \vec{B}|= 0.01 |\vec{B}|$ .

The boundary conditions for the initial magnetic field distribution
calculated with the finite element code are the following.
(i) A dipolar field along the stellar surface $r<\rs$ 
given as Dirichlet condition;
(ii) a homogeneous Neumann condition along the gap between star and disk;
(iii) a detached dipolar field along the disk (as Dirichlet condition), 
\begin{equation}
 A_{\phi}(r) = 
\left(\frac{1}{r} \frac{r^2}{(r^2 + z_{\rm D}^2)^{3/2}}\right) \tilde{A}(r) ;
\quad  {\rm with}\, \tilde{A}(r) = r^{-3/2},
\end{equation} 
and (iv) a homogeneous Neumann condition along the outer boundaries.
This implies a poloidal field inclined to the disk surface which
would support magneto-centrifugal launching of a disk wind.
The amount of 'dragging' can be defined by choosing a different function
$\tilde{A}(r)$.
The initial field is calculated with a $z$-offset, 
$z_{\rm D}^2 \ga \rs/\sqrt{2}$.
This offset avoids un-physically strong field strengths close to the
stellar surface, but leaves the field in a force-free configuration.
Again, we emphasize that such a {\em force-free} initial field is essential
in order to apply a hydro{\em static} corona as initial condition.

%---------------------------------------- Figure 1
\setlength{\unitlength}{1mm}
\begin{figure}
\parbox{50mm}
\thicklines
\epsfysize=50mm
\put(0,0){\epsffile{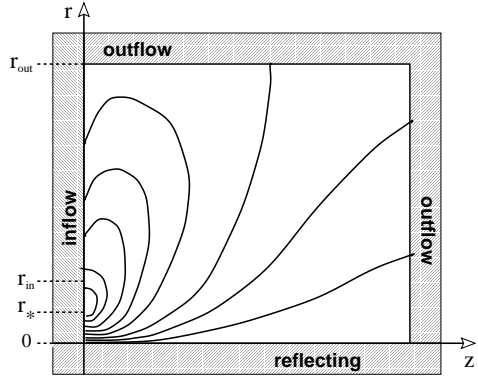}}
\caption
{Numerical model. Active region (white) and boundary region (pattern).
Star, gap and Keplerian disk are prescribed along the $r$-inflow boundary
with the stellar radius $\rs$,
inner disk radius  $r_{\rm i} \equiv 1.0$,
and maximum radius $r_{\rm out}$.
The numerical grid size is $250\times250$.
For clarification, the poloidal field lines of the initial dipole type 
magnetic field are shown as a sketch. The field smoothly continues
into the ghost zones.
}
\end{figure}

%---------------------------------------------------------------
\subsection{The boundary conditions}
The boundary condition for the poloidal magnetic field along the inflow 
boundary is fixed to the initial field.
The magnetic flux from the star and disk is conserved.
The field along the lower boundary $z=0$ corresponds to that given 
by Eq. (7).
The boundary condition for the toroidal magnetic field component
is $B_{\phi} = {\mu}_i / r $ for $r\geq r_{\rm i}$ with the parameter 
$\mu_i=B_{\phi,{\rm i}}/B_{\rm i}$.
No toroidal field is prescribed at the stellar surface.

Hydrodynamic boundary conditions are `inflow' along the $r$-axis,
`reflecting' along the symmetry axis
and `outflow' along the outer boundaries.
The inflow parameters into the corona are defined with respect to
the three different boundary regions -- star, gap and disk.
The matter is `injected' into the corona parallel to the poloidal
field lines,
$\vec{v}_{\rm inj}(r,0) = \kappa_i v_{\rm K}(r) \vec{B}_{\rm P}/B_{\rm P}$
with a density $\rho_{\rm inj}(r,0) = \eta_i\,\rho(r,0)$.
This defines the normalized mass flow rate in the disk wind, 
\begin{equation}
\dot{M}_{\rm D} = 2 \pi
\int_{r_{\rm i}}^{r_{\rm out}} 
\!\!\!\!\!\!\rho_{\rm inj,D}\,v_{\rm inj,D}\,dr =
2 \pi \eta_{\rm i,D} \kappa_{\rm i,D} 
\left( \frac{1}{r_{\rm i}} - \frac{1}{r_{\rm out}}\right)\!\!.
\end{equation}
Additionally to the disk wind boundary condition, we assume a wind 
component also from the stellar surface $r<\rs$.
The density profile of the stellar wind injection is the same
as for the disk $\rho_{\rm inj,\star}(r,0) = \eta_{i\star}\,\rho(r,0)$.
In the examples discussed in this paper, the injection velocity is chosen
as constant
$\vec{v}_{\rm inj,\star}(r,0) = 
\kappa_{\rm i,\star} v_{\rm K,i} \vec{B}_{\rm P}/B_{\rm P}$.
This gives a mass loss rate of the stellar wind component of
\begin{equation}
\dot{M}_{\star} = 2 \pi
\int_{r_0}^{r_{\star}} 
\!\!\!\!\!\rho_{\rm inj,\star}\,v_{\rm inj,\star}\,dr =
2 \pi \eta_{\rm i,\star} \kappa_{\rm i,\star} 
\left( \frac{1}{\sqrt{r_{\rm 0}}} - \frac{1}{\sqrt{r_{\star}}}\right)\!.
\end{equation}
In this case, $r_0$ is the radius of the center of the innermost 
grid element (the {\tt x2b(3)} value in the ZEUS code staggered mesh)
and is therefore biased by the numerical resolution.

This is motivated partly by numerical reasons and partly by the 
fact that {\em stellar winds} are indeed observed.
Concerning the first point,
the initial setup of a force-free magnetosphere will be distorted 
within the very first evolutionary steps giving rise to Lorentz forces.
These forces disturb the initial {\em hydrostatic} equilibrium 
resulting in a mass outflow from the regions above the star.
As a consequence, this part of the magnetosphere will be depleted of 
matter, if no additional mass inflow from the star is present.
The small density implies a strong decrease in the numerical time step
discontinuing the simulation.

On the other hand, an additional mass flow from the stellar surface is
not unlikely.
Stellar winds are common among active stars, most probably being
present also in the systems investigated in this paper.
Yet, it is not known whether stellar jets originate 
as disk winds (Pudritz \& Norman 1983) 
or as a stellar wind (Fendt et al. 1995).
Variations of the `standard' protostellar MHD jet model usually deal 
with a two-component outflow (Camenzind 1990; Shu et al. 1994). 
Model calculations of the observed emission line regions also indicate a
two-component structure (Kwan \& Tademaru 1995).
Therefore, the stellar wind boundary condition seems to be reasonable.

The ratio of the mass flow rates in the two outflow components
will definitely influence the jet structure.
Therefore, we have chosen different values for that ratio in our simulations
(Tab.\,1).
Long-term evolution runs over several hundreds of rotational periods
we have only obtained considering a rather strong stellar wind flow
which  stabilizes the region close to the rotational axis.
Since the velocity and density profiles decrease rather fast along the
disk,
the contribution to $\dot{M}_{\rm disk}$ from radii larger than
$r_{\rm out }$ is negligible.

The electro-motive force boundary condition along the inflow axis is
calculated directly from the prescribed velocity and magnetic field 
distribution,
$\vec{{\cal E}(r)} = \vec{v}(r) \times \vec{B}(r)$. 
Note that magnetic field and velocity have to be taken properly from
the staggered mesh points in order to give $\vec{{\cal E}(r)}$ as an
{\em edge-centered} property.
In the case of a stellar rotation,
$\vec{{\cal E}} \neq 0 $ for $r<\rs$.

%-------------------------------------- Table 1
%
\begin{table}
\caption[]{The table shows the parameter set varying for 
the four simulation runs S2, S4, L3, L5. Simulation
L1 is from FE99.
All the other parameters remain the same
($\beta_i = 1.0 $, $\beta_T= 0.03$, $\delta_i= 100$,
$\mu_i=-1.0$, $\rs = 0.5$). }
\label{table1}
\begin{tabular}{lllllll}
\hline\hline
\noalign{\smallskip}
 & $\kappa_{\rm i,\star}  $    & $\kappa_{\rm i,D} $ & 
   $\eta_{\rm i,\star} $ & $\eta_{\rm i,D} $ & 
   $\Omega_{\star}$ & $  \dot{M}_{\rm D}/\dot{M}_{\star}    $\\
\hline
\noalign{\smallskip}
{\bf L1} &  --      & $10^{-3}$ &  -- & $ 100 $    & -- & --\\
\noalign{\medskip}
{\bf S2} & $ 10^{-6} $ & $10^{-3}$ & $10^{3}$ &   1 & 1 & 0.5 \\
\noalign{\medskip}
{\bf S4} &  $10^{-4}$ & $10^{-3}$ & $10^3$ & $10^3$ & 1 & 2.8\\
{\bf L3} &  $10^{-4}$ & $10^{-3}$ &  200   &  200   & 1 & 1.8 \\
${\bf L5}^{*}$ &  $10^{-4}$ & $10^{-3}$ & $10^3$ &  100   & 1 & 0.2 \\
\noalign{\smallskip}
\hline\hline
\noalign{\medskip}
\end{tabular}

\noindent
{\small
*) Disk injection velocity profile is $\sim v_{\rm K}^2(r)$
}

\end{table}

%---------------

%---------------------------------------------------------------
\subsection{Numerical tests }
Before applying the ZEUS-3D code to our model we performed
various test simulations,
in particular a recalculation the OP97 2D jet simulations
(see Appendix).
Our choice of initial density distribution is stable with very good
accuracy.
Following OP97 this was tested by a run without the inflow boundary
conditions and magnetic field. 
Another signature of our proper initial magnetohydrostatic
condition is the stability of
the hydrostatic initial condition during the simulation itself.

Force-freeness of the magnetic field distribution can be tested
by calculating the $\vec{j}(r,z) \sim \nabla \times \vec{B} $ current
distribution which ideally should vanish as a consequence of the
initial condition applied.
Force-freeness can not fully be satisfied when transforming the
finite element solution to the ZEUS code initial condition,
however, an error of $0.01\,\%$ is acceptable in order to avoid
artificial effects during the first time steps until the
field distribution has evolved from its initial state.

The stability of our initial condition is demonstrated in Fig.\,A.1
showing an overlay of several initial time steps of an example
simulation without a stellar rotation presented in FE99.
The density and field distribution in the yet undisturbed regions 
perfectly match during the first decades of evolution ($t<75$).

%---------------------------------------- Figure 2
\setlength{\unitlength}{1mm}
\begin{figure*}
\parbox{250mm}
\thicklines
\epsfxsize=40mm
 \put(5,95){\epsffile{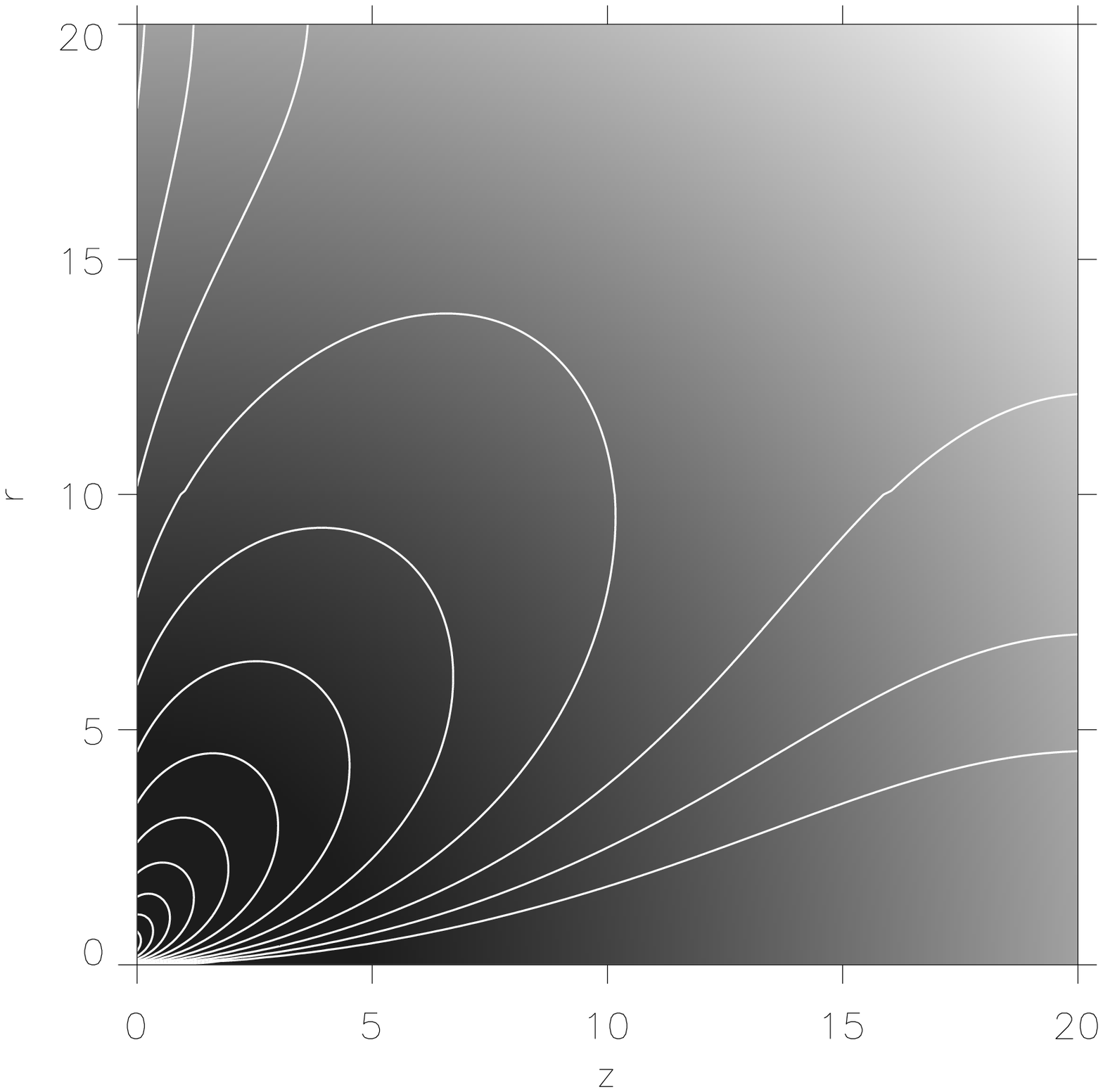}}
 \put(50,95){\epsffile{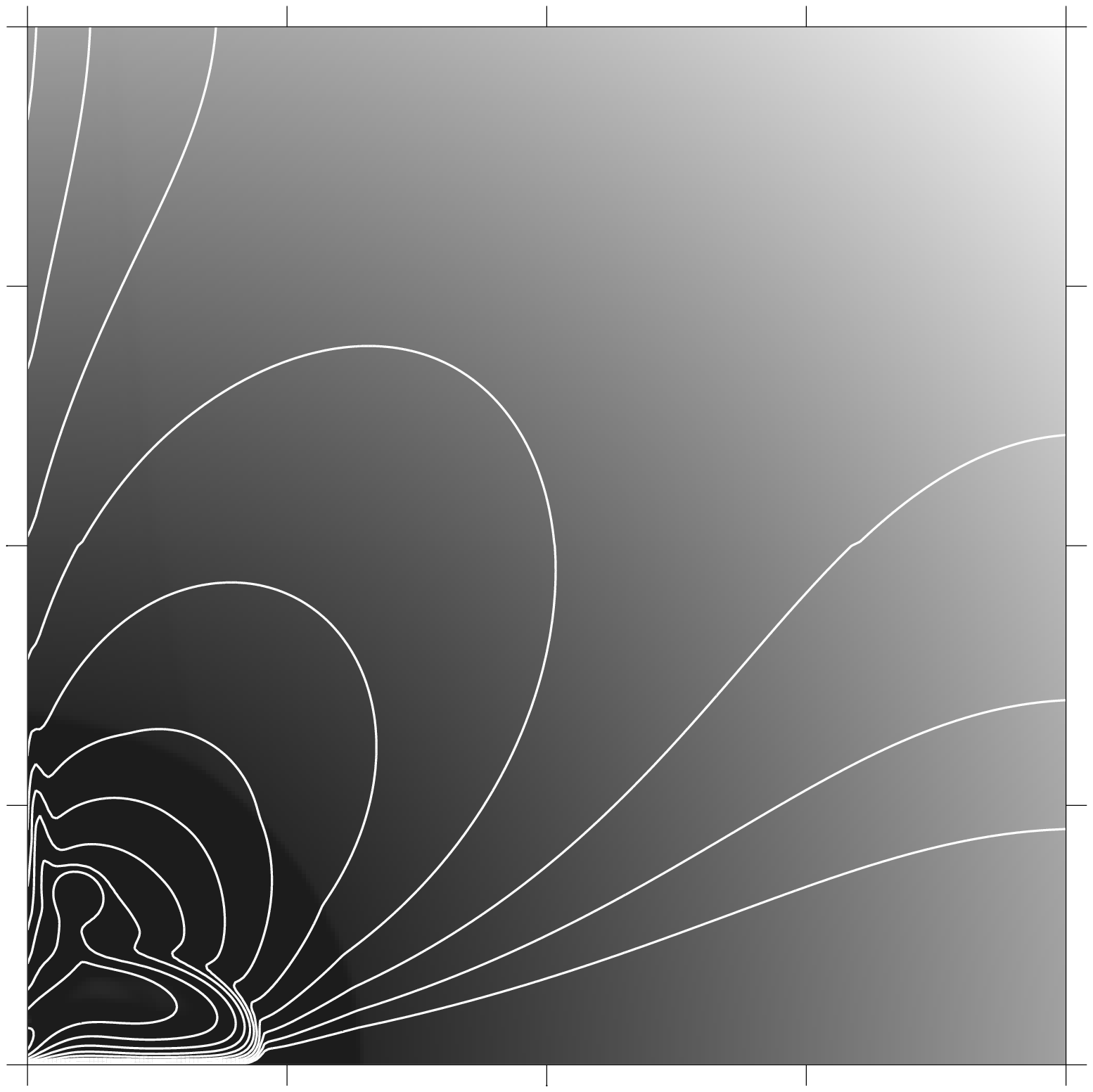}}
 \put(95,95){\epsffile{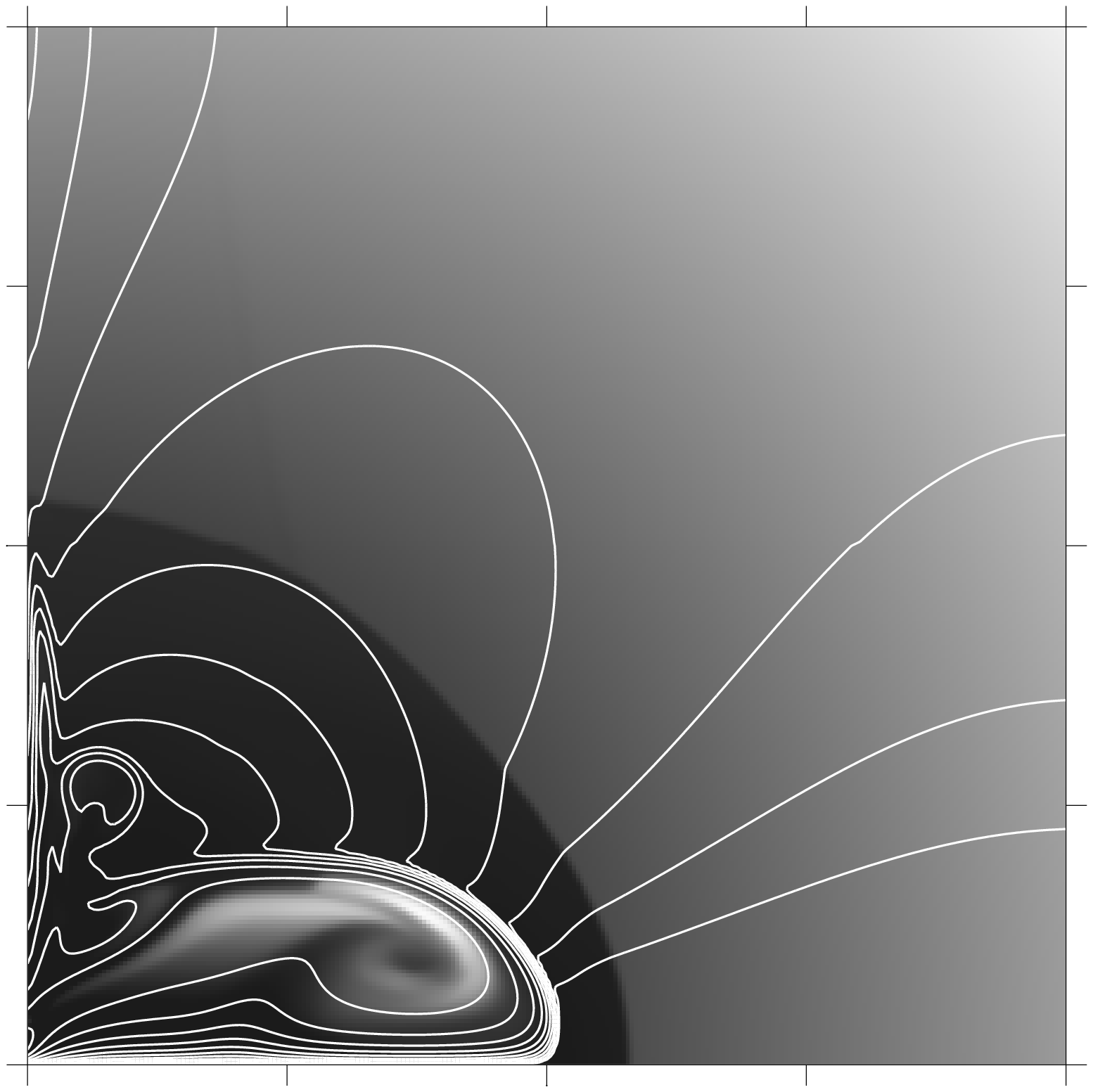}}
 \put(140,95){\epsffile{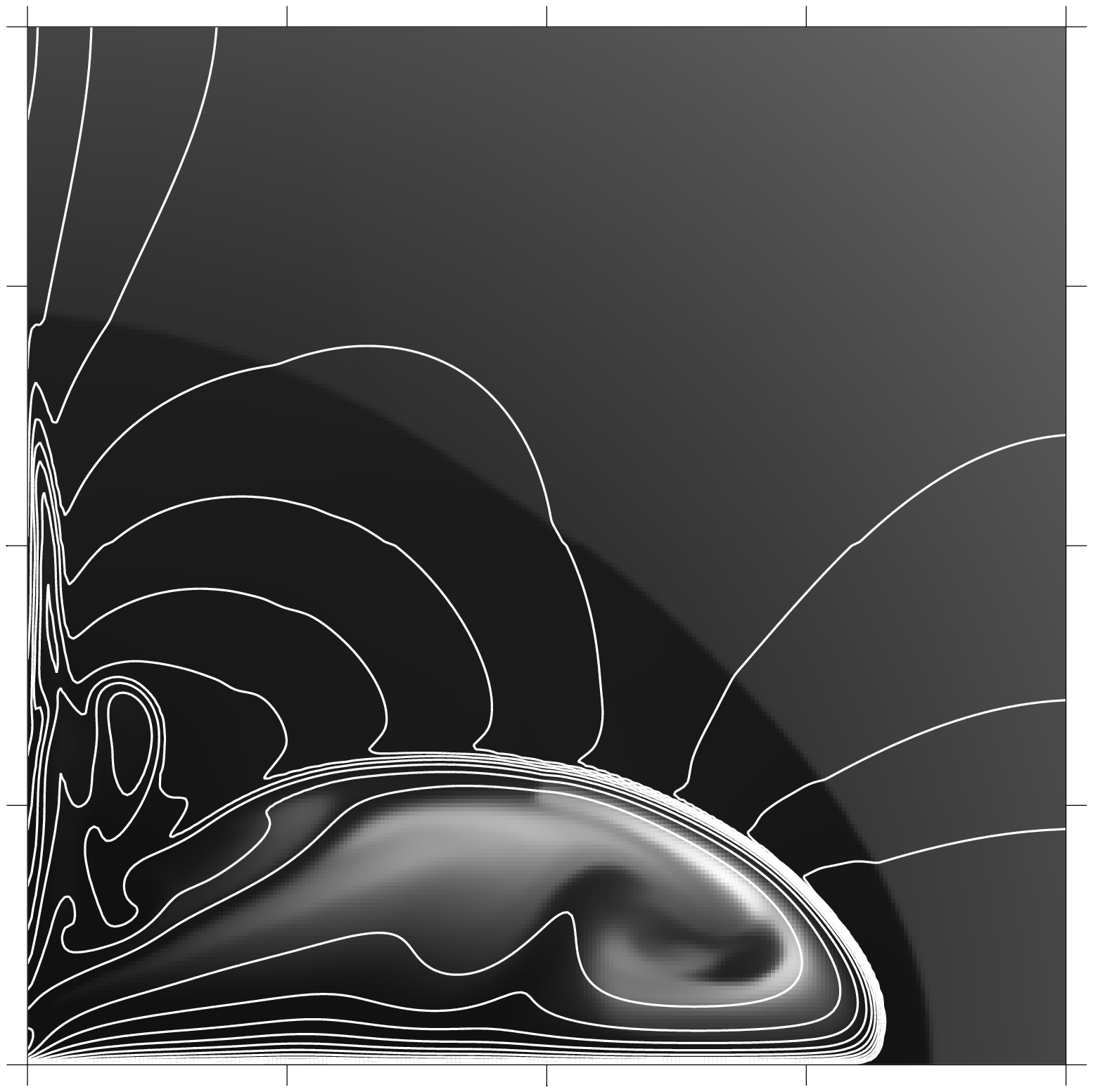}}
%
%\put(5,47){\epsffile{9822f2e.ps}}
%\put(50,47){\epsffile{9822f2f.ps}}
%\put(95,47){\epsffile{9822f2g.ps}}
%\put(140,47){\epsffile{9822f2h.ps}}
%
 \put(5,2){\epsffile{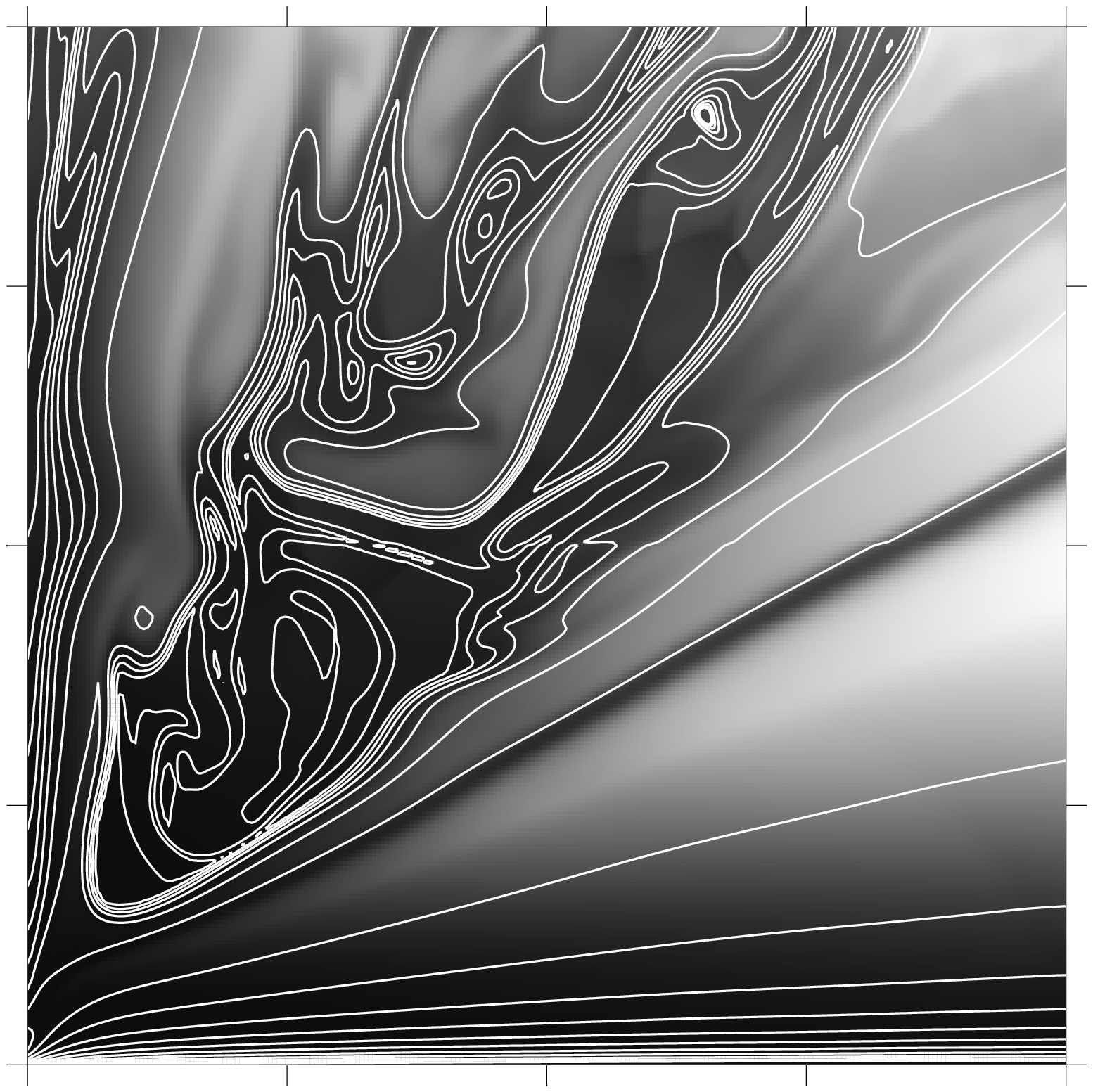}}
 \put(50,2){\epsffile{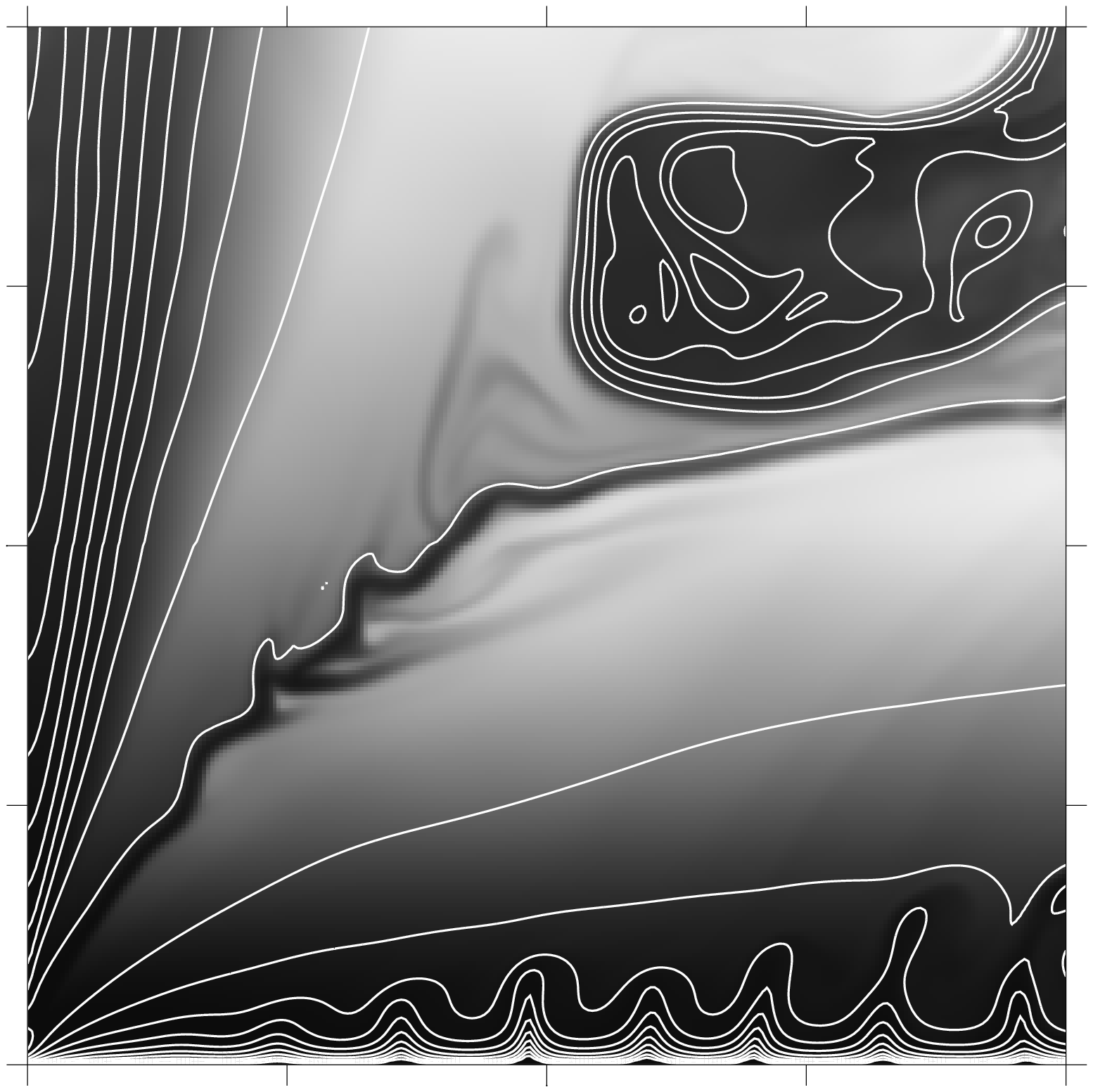}}
 \put(95,2){\epsffile{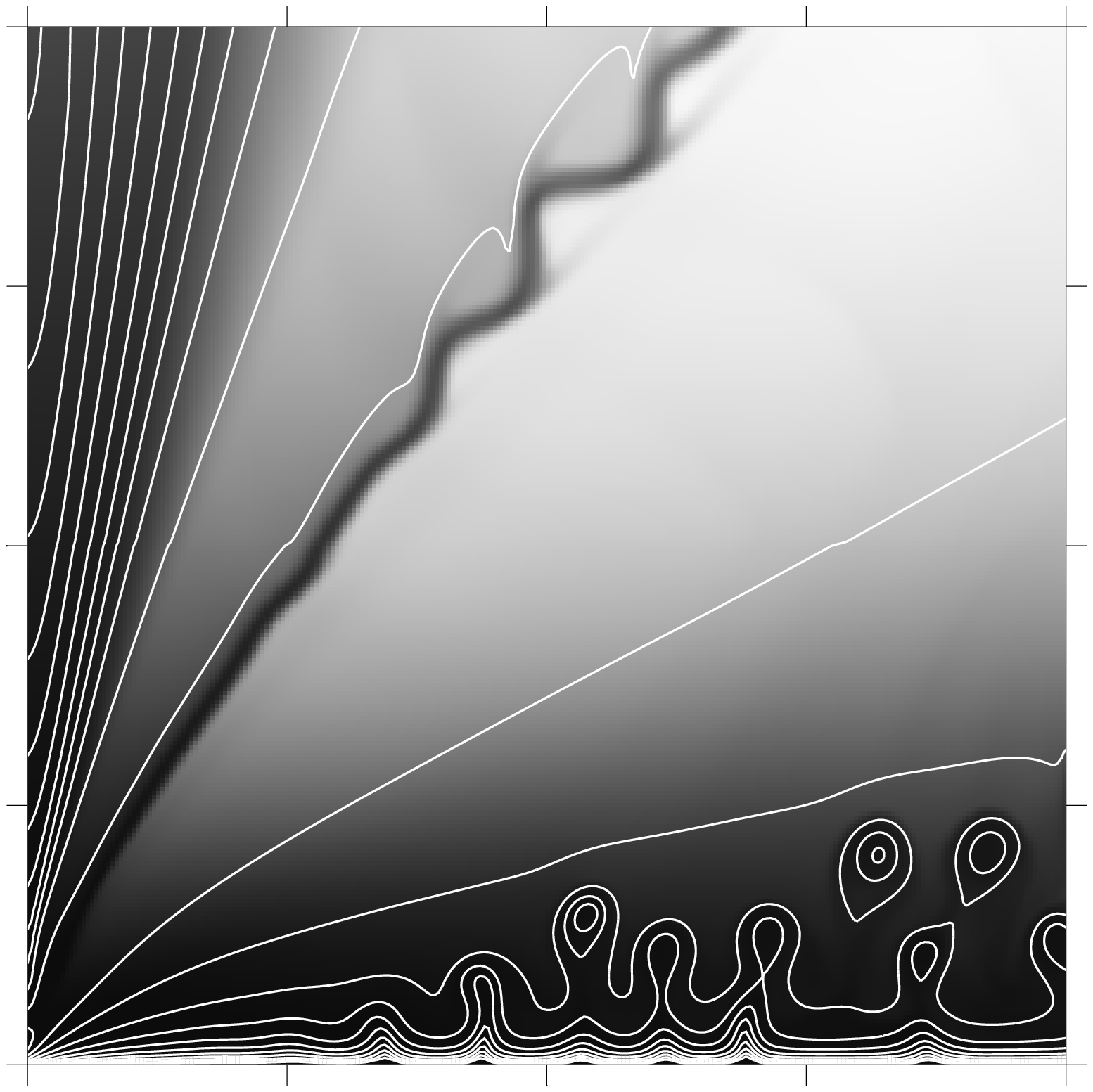}}
\epsfxsize=40mm
 \put(140,2){\epsffile{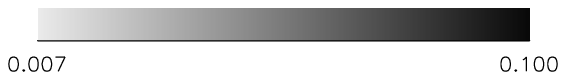}}
\caption
{Simulation S2 in a box of $20\times 20\,r_{\rm i}$.
Shown are density (grey scale) and poloidal field lines (contour lines)
for 
$t = 0, 10, 20, 30, 50, 100, 200, 300, 500, 1000, 2500 $
(from {\it left} to {\it right} and {\it top} to {\it bottom}).
The density at the inner disk radius is 
$\rho_{\rm i} = 1.0$.
The legend shows the density limits used for the color coding
(which itself uses the inverse density profile).
The stellar radius is $\rs = 0.5\,r_{\rm i}$.
}
\end{figure*}

%---------------------------------------- Figure 3
\setlength{\unitlength}{1mm}
\begin{figure*}
\parbox{100mm}
\thicklines
\epsfxsize=40mm
%\put(5,45){\epsffile{9822f3a.ps}}
%
\epsfxsize=35mm
%
%\put(0,5){\epsffile{9822f3b.ps}}
%\put(36,5){\epsffile{9822f3c.ps}}
%\put(72,5){\epsffile{9822f3d.ps}}
%\put(108,5){\epsffile{9822f3e.ps}}
%\put(144,5){\epsffile{9822f3f.ps}}
%
\caption
{Simulation S4 in a box of $10\times 10\,r_{\rm i}$.
Density and poloidal field lines.
Notation equivalent to Fig.\,2. 
Time steps 
$t = 0, 10, 40, 400, 600 t_i$. 
}
\end{figure*}

%---------------------------------------- Figure 4
\setlength{\unitlength}{1mm}
\begin{figure*}
\parbox{100mm}
\thicklines
\epsfxsize=40mm
%\put(5,90){\epsffile{9822f4a.ps}}
%
\epsfxsize=35mm
%
%\put(0,50){\epsffile{9822f4b.ps}}
%\put(36,50){\epsffile{9822f4c.ps}}
%\put(72,50){\epsffile{9822f4d.ps}}
%\put(108,50){\epsffile{9822f4e.ps}}
%\put(144,50){\epsffile{9822f4f.ps}}
%
%\put(0,5){\epsffile{9822f4g.ps}}
%\put(36,5){\epsffile{9822f4h.ps}}
%\put(72,5){\epsffile{9822f4i.ps}}
%\put(108,5){\epsffile{9822f4j.ps}}
%\put(144,5){\epsffile{9822f4k.ps}}
%
%
\caption
{High resolution simulation L3 in a box of $5\times 5\,r_{\rm i}$.
{\it Above} Density and poloidal field lines.
Notation equivalent to Fig.\,2. 
{\it Below} Contour plots of the toroidal magnetic field strength.
The toroidal field is positive (negative) outside (inside) 
the neutral line.
Time steps are
$t = 0, 10, 20, 80, 160 $ 
(from {\it left } to {\it right}).
}
\end{figure*}

%---------------------------------------------------------------
\section{Results and discussion}
%---------------------------------------------------------------
%
In the following we discuss the results of four example
simulations denoted by S2, S4, L3, L5 (see Tab.\,1).
All the simulations presented in this paper consider a rotating 
star at the center.
The stellar rotational period is chosen as 
$\Omega_{\star} = (v_{K,i}/R_i) = 1$,
with a magnetospheric co-rotation radius located at the inner disk
radius.
The evolution of a non-rotating central star is discussed in FE99,
although it was not possible to perform the simulation as long as the
examples presented here.

As a general behavior, the initial dipole type structure of the
magnetic field disappears on spatial scales larger than the
inner disk radius and a two component wind structure -- a disk
wind and a stellar wind -- evolves.
Our main result is the finding of a {\em quasi-stationary} final state of
a spherically radial mass outflow evolving from the initial
dipole type magnetic field on the very extended time evolution.
For the boundary conditions applied the calculated flow structure
show only few indication for collimation.

The general features in the evolution of the system are independent from a
variation of the field strength.
For strong fields, the evolution is faster.
Thus, for simulations which are limited in time due to numerical problems,
a decrease in the field strength would not help.
Although the numerical life time of the simulation would be accordingly 
longer, the result for the final time step will be the same.

%---------------------------------------------------------------
\subsection{Four simulations of the long-term magnetospheric evolution 
            -- an overview}
The four example simulations basically differ in the mass flow rates
from disk and star and the size of the physical domain investigated
(Tab.\,1).
All other parameters,
$\beta_i = 1.0 $, $\delta_i= 100$, $\mu=-1.0$, $\rs = 0.5$,
$\gamma = 5/3$,
and the numerical mesh of $250\times 250$ grid elements remain
the same.
The first simulation (solution {\bf S2}) considers a rectangular box
of physical size of 20x20 inner disk radii (Fig.\,2, Fig.\,5, Fig.\,7).
The stellar wind mass flow rate is comparatively large, 
$\dot{M}_{\star}/ \dot{M}_{\rm D} = 2$. 
The stellar wind injection velocity is very low in order to not disturb 
a possible weak wind solution already by the boundary condition.
The disk wind boundary condition is the same as in OP97 and FE99.
Due to the relatively large physical size of the computational domain
the stability of the initial condition can be observed for several
decades of rotational periods. 
The torsional Alfv\'en waves leave the grid after about $t= 40$.
The second simulation (solution {\bf S4}) considers a rectangular box
of physical size of 10x10 inner disk radii (Fig.\,3, Fig.\,5, Fig.\,7).
Now the total mass flow is dominated by the disk wind .
This choice of parameters can be directly seen from the simulation
by comparing the size of axial flow and the 'bubble' evolving from the
disk flow. Clearly, the disk flow is more prominent.
The third (solution {\bf L3}) considers a rectangular box
of physical size of only 5x5 inner disk radii (Fig.\,4, Fig.\,6).
Similar to solution S4, the total mass flow is dominated by the disk wind.
This high resolution simulation zooms into the innermost region around
the star.
In particular, the neutral line is clearly resolved.
We finally discuss another example ({\bf L5}) in which the stellar
wind is dominating the disk wind.
This simulation perfectly evolves into final stationary
state.
For L5 the velocity injection profile is chosen differently from the
examples discussed above in order to increase the disk flow magnetization.
All parameter runs show a similar gross behavior indicating that our 
run S2 with lowest resolution is sufficient in order to investigate
the main features of the flow evolution.
In general, a high stellar wind mass loss rate will stabilize
the outflow.

%---------------------------------------------------------------
\subsection{The first evolutionary stages}
During the first stages of the long-term evolution the magnetospheric
structure is characterized by the following main features:
The {\em winding-up} of the dipolar poloidal field,
the formation of a {\em neutral field line},
a {\em transient axial jet} feature,
a two component outflow consisting of
a {\em stellar wind} and a {\em disk wind}.
 
%---------------------------------------------------------------
\subsubsection{Winding-up of the poloidal field}
The winding-up process of poloidal magnetic field due to differential
rotation between star and disk and the static initial corona
induces a toroidal field (Fig.\,4) 
with a positive sign along field lines 
located outside the slowly emerging neutral field line.
Inside the neutral field line $B_{\phi}$ has negative sign.
This is in difference to OP97 and other simulations assuming a
monotonously distributed initial field.

Torsional Alfv\'en waves propagate outwards distorting the 
initial field structure.
After about $t=40$ these waves reach the outer boundary (Fig.\,2).
The region beyond the wave front remains completely undisturbed.
The region between the Alfv\'en wave front and the flow bow shock 
is adjusted to a new equilibrium and also remains in equilibrium
until it is reached by the generated outflow 
(
See the density contour lines close to the disk in Fig.\,A2.
The grey scale density plots cannot show this feature
).
Within this Alfv\'en wave front the magnetic field becomes distorted
from its initial force-free state
(compare the field line structure of the closed loops in Fig.\,2 for
$t \leq 30$).
The distortion of the force-free field due to propagating Alfv\'en waves
results in Lorentz forces initiating an axial `jet' feature close to 
the axis. As we will see later, this axial jet, however, is a
transient feature.

The winding-up of poloidal magnetic field seems to be similar to the
effect proposed by Lovelace et al. (1995).
However, in our case, this process is initiated by the differential 
rotation between star and hydrostatic corona.
Only later, the wound-up toroidal field is maintained by both,
differential rotation between star and disk and the inertia of the 
outflow.

%---------------------------------------------------------------
\subsubsection{A neutral field line dividing stellar and disk field}
The wound-up magnetic field lines stretch forming a neutral field
line of vanishing field strength.
The matter around this field line is distributed in a layer of low
density (Fig.\,4).
Around this layer an expanding `bubble' is formed due to the
additional magnetic pressure due to toroidal fields
which disrupts the initial dipolar field structure (Fig.\,3).
When the bubble has left the grid, the field lines which are 
separated by the the neutral line remain disconnected.
This is due to the differential rotation between star and disk. 
The actual appearance of the axial jet and the low density bubble
depends mainly on the mass flow rates from disk and star.
The bubble is most prominent in simulation S4 where the disk mass
flow rate is largest.

%---------------------------------------------------------------
\subsubsection{A transient axial jet feature}
In the beginning of the simulations a jet feature evolves along
the rotational axis.
Its pattern velocity is about $0.2\,v_{\rm K,i}$ (S2) or
$0.3\,v_{\rm K,i}$ (L3).
Such an axial jet is known as a characteristic result of MHD simulations
performed in the recent literature 
(Hayashi et al. 1997; Goodson et al. 1997, 1999; Goodson \& Winglee 1999;
Kudoh et al. 1998).
It is often claimed that this feature is connected to the real 
(protostellar) jets observed on the AU-scale.
Apart from the fact that the spatial dimension and velocity 
do not fit with the observations, 
we will see that the formation of this feature is a result of the
adjustment process of the initially hydrostatic state to a new dynamic
equilibrium and will disappear on the long-term evolution.

Winding-up of the initially force-free magnetic field by differential
rotation during the first time steps immediately leads to a not
force-free field configuration.
The resulting magnetic forces accelerate the material of the initially
hydrostatic corona forming an axial jet.
This process works as long as initially distributed coronal matter
is present at this location.
As time develops the jet feature becomes weaker and weaker.
Disk wind and the stellar wind become the dominant flow pattern
and the axial jet dies out after about $t>100$ (Fig.\,2). 
In comparison, in simulation L1 (FE99), where a stellar wind is absent,
the coronal density along the axis decreases until it is below
the numerically critical value and the simulation stops.

The intermittent character of the axial jet flow is best seen in 
the velocity structure (Fig.\,5).
The velocity vectors of the axial flow are largest during
the first time steps.
However, after sweeping-out the initial corona, a weak wind flow 
from the stellar surface succeeds the jet.

%---------------------------------------------------------------
\subsubsection{The disk wind}
The disk wind accelerates within tens of grid elements from its
low injection speed to fractions of the Keplerian speed.
The acceleration mechanism is mainly due to the centrifugal 
force on the disk matter reaching the non-rotating corona, 
where the gravitational force is balanced by the 
pressure and not by a centrifugal force as in the disk.  
The flow is already super Alfv\'enic due to the weak dipolar field
strength in the disk.
Thus, magneto-centrifugal acceleration along the inclined 
dipole type field lines of the initial magnetic field is
{\em not} the acceleration mechanism.
More above the disk also the Lorentz force 
along the field contributes to the acceleration (see Sect.\,4.4.3.).
The inclination angle between field lines (equivalent to the outflow
direction) and the disk depends on the mass flow rate. 
For the parameter range investigated we see no indication for
a disk wind collimation because the Lorentz force points away
from the axis (see below).

%---------------------------------------------------------------
\subsubsection{The stellar wind}
The rotating stellar magnetosphere generates a stellar wind.
Due to the strong magnetic field close to the star the 
flow starts sub-Alfv\'enic.
It is initially magneto-centrifugally driven with a roughly 
spherical Alfv\'en surface located at $1.5\ri$ (L3) or
closer (S2, S4, L5) to the stellar surface.
The most dominant flow pattern of the stellar wind is in the
part with the widest opening angle (Fig.\,5). Although the 
Lorentz force points radially inwards no collimation is observed 
because of a strong pressure gradient.
Depending on the mass loss rates the stellar wind evolves faster
or slower than the disk wind (Fig.\,5).

%---------------------------------------- Figure 5
\setlength{\unitlength}{1mm}
\begin{figure}
\parbox{200mm}
\thicklines
\epsfysize=40mm
\epsfxsize=40mm
%
%\put(0,160){\epsffile{9822f5a.ps}}
%\put(40,160){\epsffile{9822f5b.ps}}
%
%\put(0,120){\epsffile{9822f5c.ps}}
%\put(40,120){\epsffile{9822f5d.ps}}
%
%\put(0,80){\epsffile{9822f5e.ps}}
%\put(40,80){\epsffile{9822f5f.ps}}
%
%\put(0,40){\epsffile{9822f5g.ps}}
%\put(40,40){\epsffile{9822f5h.ps}}
%
%\put(0,0){\epsffile{9822f5i.ps}}
%\put(40,0){\epsffile{9822f5j.ps}}
%
\caption
{Evolution of the poloidal velocity in the simulations 
S2 ({\it left}) and 
S4 ({\it right}). 
Time steps (from {\it top} to {\it bottom}).
$t = 10, 20, 100, 500, 2500 $ (S2) and
$t = 5, 10, 20, 100, 600 $ (S4).
Vectors scale only within each frame.
}
\end{figure}

%---------------------------------------- Figure 6
\setlength{\unitlength}{1mm}
\begin{figure}
\parbox{190mm}
\thicklines
\epsfysize=49mm
\epsfxsize=40mm
%
%\put(0,160){\epsffile{9822f6a.ps}}
%\put(45,160){\epsffile{9822f6b.ps}}
%
%\put(0,120){\epsffile{9822f6c.ps}}
%\put(45,120){\epsffile{9822f6d.ps}}
%
%\put(0,80){\epsffile{9822f6e.ps}}
%\put(45,80){\epsffile{9822f6f.ps}}
%
%\put(0,40){\epsffile{9822f6g.ps}}
%\put(45,40){\epsffile{9822f6h.ps}}
%
%\put(0,0){\epsffile{9822f6i.ps}}
%\put(45,0){\epsffile{9822f6j.ps}}
%
\caption
{Highly time-resolved evolution of simulation L3. 
Poloidal magnetic field lines ({\it left}) and
density contours ({\it right}).
Time steps 
$t = 200, 201, 202, 203, 204 $ 
(from {\it top} to {\it bottom}).
}
\end{figure}

%---------------------------------------------------------------
\subsection{The long-term evolution}
The long-term evolution of the flows depends critically
on the choice of inflow boundary conditions into the corona.
The total mass flow rate into the corona determines how fast 
the flow will establish a (quasi-)stationary state.
The stellar wind - disk wind mass flow ratio determines
(i) the opening angle of the outflow, 
(ii) the opening angle of the cone of the neutral line which is the 
boundary layer between the stellar wind and the disk wind,
and (iii) the stationarity of the axial flow
(see Tab.\,2).
In general the outflow undergoes a highly time-variable and turbulent 
evolution.
However, after relaxation of the MHD system from the initial
magneto hydrostatic state into the new {\em dynamical}
equilibrium, 
we observe an outflow from disk and star distributed smoothly over the
whole hemisphere and moving predominantly in spherically radial
direction.

In simulation S2 the flow structure is highly time-variable over
many hundreds of periods.
The intermediate region between the two components 
-- stellar wind and disk wind --
is characterized by turbulent motions of very low velocity.
But also the disk wind seems to be unstable.
During the intermediate time evolution only the flow pattern along 
the $z$-axis relaxes.
\clearpage
Compared to the evolution of the stellar wind, the disk wind needs 
definitely more time to establish a stationary structure.
However, after all these turbulent evolutionary steps,
with few changes in the general appearance of the flow pattern
over hundreds of rotational periods,
{\em after about 2000-2500 rotations a quasi-stationary
outflow has been established} over the whole grid 
(Fig.\,2). 
Only the region around the neutral line dividing stellar and disk
wind and the region along the $z$-axis is subject to small scale 
instabilities.
Interestingly, the flow along the symmetry axis which shows a stable
behavior during the intermediate time evolution, finally becomes
unstable.
A conical flow consistent of a knotty structure evolves
with a full opening angle of $30\degr$.
At this time the opening angle of the neutral layer cone dividing
stellar wind from disk wind has been increased compared to the
intermediate time steps.
This de-collimation of the outer flow causes a de-stabilization of
the axial flow.
In this sense, the stellar wind flow is stabilized by the ambient
(disk wind) pressure.
The formation of knots and instabilities along the
symmetry axis depends on the stellar wind mass flow rate (see below).
The full opening angle of `knot flow` is about 30\degr to 50\degr.
We emphasize that due to the knot size and knot spacing,
these features are not correlated to the observed knots of protostellar
jets. It seems more related to QPO's observed in X-ray binaries.
The knot velocity is about 10\% of the Keplerian speed at the disk inner
radius.

In simulation S4 the quasi-stationary state is reached
earlier after about 200 inner disk rotations according
to the smaller box size (Fig.\,3).
The region around the neutral field line is resolved better.
We observe the formation of a wavy structure with an amplitude
of $0.5\,\ri$ and a wavelength of $\ri$.
These `waves' travel outwards and leave the grid.
The wave generation is somewhat arbitrary in time.
Whereas the region enclosing the neutral layer first seems to have reached a
steady state at about $t=100$, 
at $t=600$ the wave structure evolves again.
This is related to the evolution along the symmetry axis.
Here, in the contrary, at $t=100$ the simulation still showed a wavy pattern,
whereas at $t=600$ a smooth steady state has evolved.
The flow structure at $t=400$ seems to be completely smooth and stable.
The long-term evolution shows that this stability is in fact a transient
feature as far as the neutral layer region is concerned (see below).

Simulation L3 shows that the neutral line has a complex structure.
Two current sheaths emerge, one from the stellar radius, the other
from the inner disk radius (Fig.\,4).
These are indicated by the density and poloidal magnetic field
`islands' emitted along this field line 
(Fig.\,4, time step 80 and 160).
Figure 6 shows the evolution of the solution L3 with a high time
resolution $\Delta t = 1$ after $t=200$.
At this time the simulation has not yet evolved into a 
quasi-stationary state.
It can be seen how plasmoids are formed and move outwards
along the neutral line.
Most probably, this would be a region of on-going reconnection 
processes.
A similar behavior was found first by Hayashi et al. (1997)
including also magnetic diffusion in their treatment.
However, the simulation lasted only for one inner disk rotational
period (with a star at rest).
Our long-term simulations show that the formation of such plasmoids 
will continue.
We do not believe that the lack of diffusion in our treatment
is a serious problem concerning this point
because the 
time scale given by the plasmoid velocity is
smaller than the time scale given by magnetic diffusion.

Figure 5 shows the poloidal velocity vectors of the simulations S2 and
S4 at selected time steps (compare to Fig.\,3 and 4).
The general feature is that the axial jet feature seen in the first
time steps disappears.
The outflow exhibits a two-component structure.
Depending on the inflow density profile the `asymptotic' (i.e. close
to the grid boundaries) velocity profile changes slightly.
High velocities (larger than $v_{\rm K,i}$) are only observed far from
the axis.
For solution S2 we obtain an asymptotic speed of 
$1.5\,v_{\rm K,i}$ for both components with a velocity profile
decreasing across the neutral field line.
The flow velocity along the axis is $0.2\,v_{\rm K,i}$.
For solution S4 the asymptotic speed of $1.5\,v_{\rm K,i}$ 
for the stellar wind component and with a velocity profile
increasing across the neutral field line $2\, v_{\rm K,i}$.
The latter is due to the low density at this point.
The disk wind has not yet accelerated to high velocities.
This is due to the wide opening angle of the stellar wind cone
and the small physical grid size.
The flow velocity along the axis about $0.5\, v_{\rm K,i}$.
For solution L5 the `asymptotic' speed of the stellar wind is 
low and $\leq v_{\rm K,i}$.
The velocity profile is in general similar to that of solution S4.

The duration of our simulation runs (L3, L5, S2, S4) is not limited by
numerical reasons.
They have been terminated when the flow evolution has reached a {\em quasi}
stationary state.
For simulation S2 this means that after about 2500 rotations the main pattern
in the flow evolution does not change anymore.
In particular, the outer part of the wind flow does not vary in time,
while knots are still generated along the axis.
For simulation S4, after 400 time steps most of the flow region is in a
stationary state.
Only the wavy structure along the neutral field line, which is also 
connected to the formation of knots, continuously evolves and disappears.
This structure is {\em not} a dangerous instability for the 
outflow (see also Fig.\,7).
For solution L3, the simulation has been terminated when parts of the 
flow (the outer disk wind flow) had reached a stationary state. 
The inner solution close to the axis does not reach such a stationary 
state.
In this case, our intention is to investigate the neutral line with better
spatial resolution.
Simulation L5 was terminated some time after the stationary state
has been reached for the whole outflow.

%---------------------------------------------------------------
\subsection{A stationary final state: a radial outflow 
evolved from an initially dipolar magnetosphere}
The main result of our simulations is that the initial dipolar
magnetosphere evolves into a spherically radial outflow consisting
of two components.
Depending on the inflow parameters (mass flow rates, magnetic field 
strength) our simulations reach a quasi-stationary state.
A weak non-stationarity may be present along the neutral field
line, which is dividing the stellar wind from the disk wind.
Also, for a weak stellar wind flow a turbulent flow pattern
may evolve along the axis.
Such outflows we call {\em quasi-stationary}, if the main flow
pattern does not change in time.
The disk wind and the outer cone of the stellar wind reach a kind
of stationary state, where the density profile and field line
structure remain almost constant in time.

For the S2 solution the half opening angle of stellar wind cone 
is about $55\degr$.
This angle remarkably changes during the flow evolution. During the
initial evolutionary decades the turbulent region between the stellar
wind and the disk wind collimated the stellar wind to a narrower cone.
Clearly, such a neutral line is a rather unstable situation.
Reconnection will most probably occur which we cannot properly
treat with our ideal MHD approach.
Increasing the numerical resolution (simulation L3) shows the emission 
of plasmoids along the neutral line (see below).
Figure 7 shows an overlay of three time steps of solution S2 at
$t= 2600, 2650, 2700$ clearly indicating the stationary character
of disk wind and most parts of the stellar wind
together with the non-stationary axial flow 
and the small-scale wave pattern along the neutral field line.

%-------------------------------------- Table 2
%
\begin{table}
\caption[]{Terminal poloidal velocity of the wind components 
from the disk $v_{\rm p, disk}^{\rm max} $, 
from the star $v_{\rm p,star}^{\rm max} $, 
along the axis $v_{\rm p, axial} $,
and from the gap $v_{\rm p, gap}^{\rm max} $,
time when stationary state has been reached $t_{\rm s}$,
and the inclination angle between disk and neutral layer $\alpha$
for the different simulations.}
\label{table2}
\begin{tabular}{lllllll}
\hline\hline
\noalign{\smallskip}
 & $v_{\rm p, disk}^{\rm max}  $ & $v_{\rm p,star}^{\rm max} $ & 
   $v_{\rm p, axial} $ & $v_{\rm p, gap}^{\rm max} $ & 
   $t_{\rm s}$ & $\alpha $ \\
\hline
\noalign{\smallskip}
{\bf L1} & 1.0 & ? & ?  & ? & ? & ? \\
\noalign{\medskip}
{\bf S2} & 1.0 & 1.5 & 0.2 & 1.5 & 2500   & $35\degr$ \\
\noalign{\medskip}
{\bf S4} & 1.5 & 2.1 & 0.6 & 1.5 & 400 & $60\degr$  \\
{\bf L3} & 1.0 & 1.3 & 0.2 & 1.7 & ?   & $50\degr$  \\
{\bf L5} & 0.9 & 1.0 & 0.2 & 3.0 & 150 & $35\degr$  \\
\noalign{\smallskip}
\hline\hline
\noalign{\medskip}
\end{tabular}

\end{table}

In the case of solution S4 the quasi-stationary state is reached
earlier after about $t=400$ (Fig.\,3).
The comparatively large stellar wind mass flow rate stabilizes
the flow along the axis and no turbulent pattern evolves.
On the other hand, due to the low disk wind mass flow rate 
the half opening angle of the neutral line cone is smaller.
It is $35\degr$ compared to $55\degr$ for solution S2.
Again, Fig.\,7 demonstrates the stationarity of the simulation
with an overlay of three time steps $t= 575, 600, 625$.
In comparison with solution S2 now the whole flow pattern 
is stationary.
In particular the region along the axis remains stable.
The only time-dependent feature is the neutral line exhibiting
a slowly variable wave structure.

%---------------------------------------- Figure 7
\setlength{\unitlength}{1mm}
\begin{figure*}
\parbox{100mm}
\thicklines
\epsfysize=86mm
\epsfxsize=70mm
 \put(0,75){\epsffile{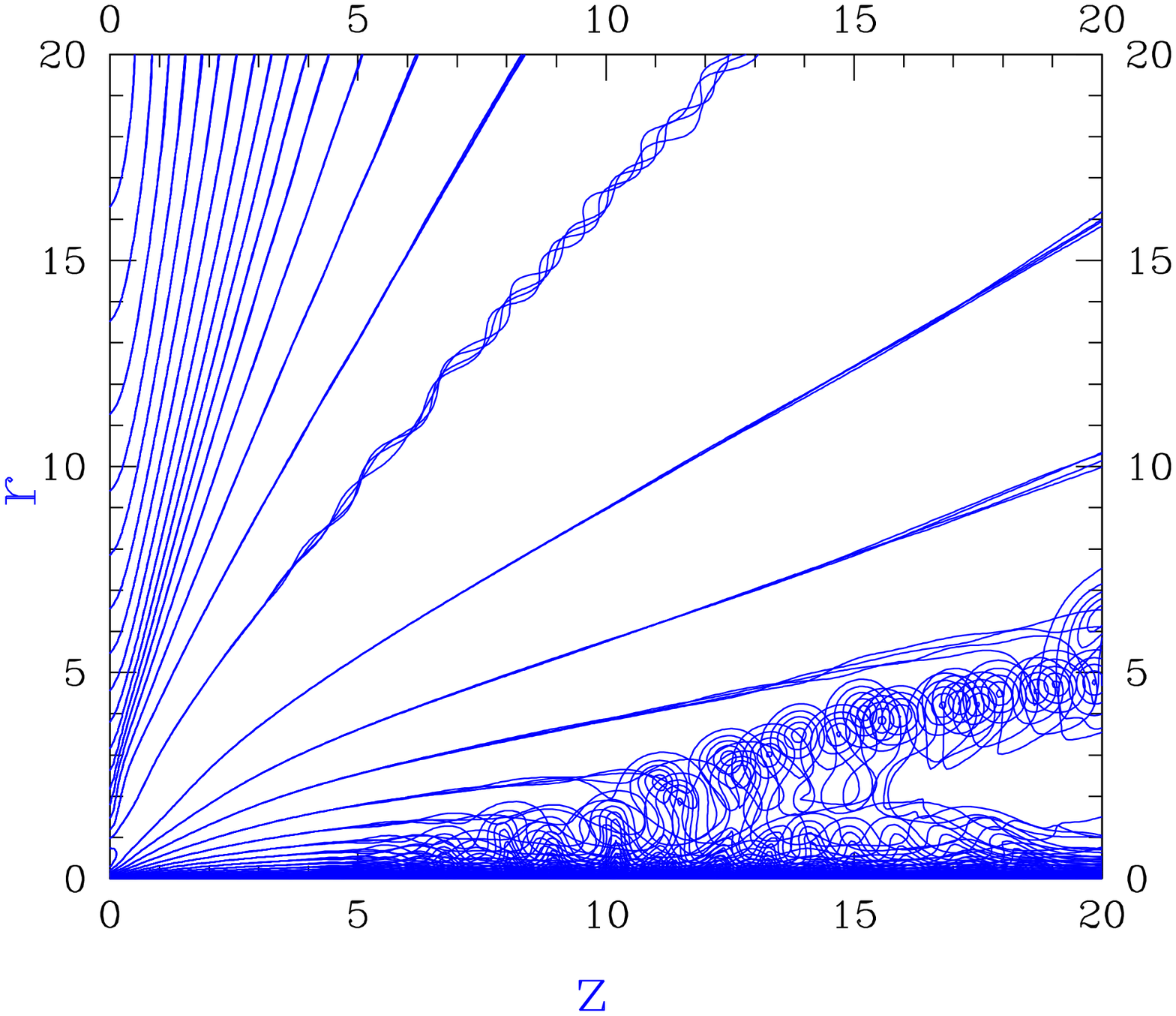}}
 \put(80,75){\epsffile{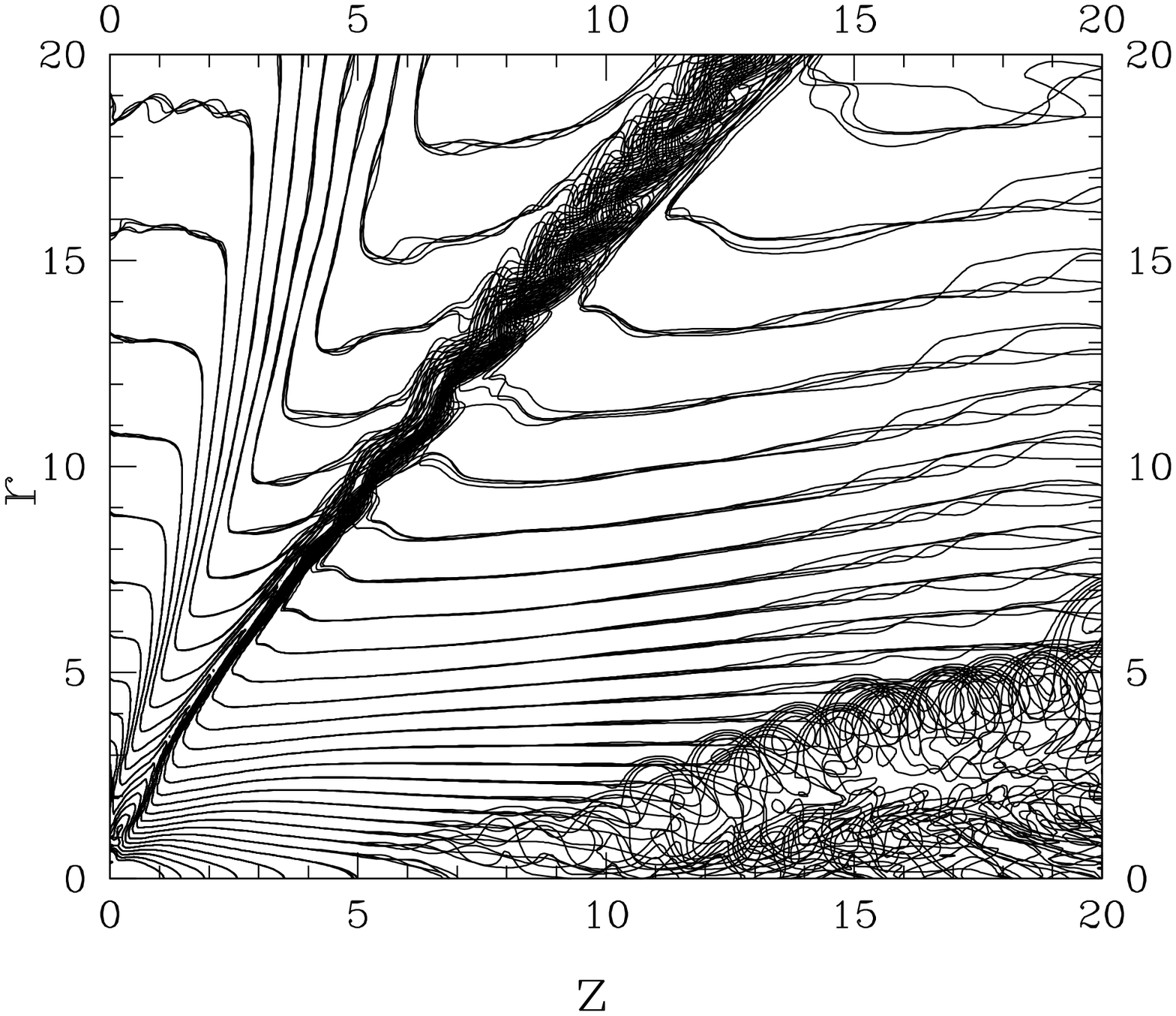}}
%\put(0,0){\epsffile{9822f7c.ps}}
%\put(80,0){\epsffile{9822f7d.ps}}
%
\caption
{Evolution of simulation S2 (upper) and S4 (lower) on the very
long time scale.
Shown are overlays of the 
poloidal field lines ({\it left}) and
density contours ({\it right}).
Three time steps are superposed, 
$\tau = 2600, 2650, 2700$ (S2) and
$\tau = 575, 600, 625$ (S4).
}
\end{figure*}
%-------------------------------------------------

Increasing (i) the total mass flow rate and (ii) also the 
stellar wind to disk wind mass flow ratio  
will result in an almost perfectly
stationary flow (solution L5; see below).
Having a similar mass flow ratio, also the opening angle is similar
to solution S2.
The stationarity of solution L5 will be investigated in more detail 
below.

The blobs (or rather tori) generated in simulation S2 
move with pattern speed of
about 0.1 the Keplerian speed at $r_i$.
Their size is about the inner disk radius but depends from the mass
flow ratio and the numerical resolution.
We emphasize that due to the knot size and time scale of knot formation 
in our simulation,
their connection with the jet knots observed in protostellar jets
on the large scale distance of tens of AU is questionable.
This statement also holds for comparable structures observed in similar
simulations presented in the literature
(Ouyed \& Pudritz 1997b,
Goodson et al. 1997, 1999; Goodson \& Winglee 1999).

On the other hand, from the unstable character of the axial flow
together with the lack of collimation we may conclude that the
model configuration investigated in our paper is unlikely to
produce collimated jets.
Furthermore, we hypothesize that this behavior may be the one of the 
reasons why highly magnetized star-disk systems -- containing magnetic
white dwarfs or neutron stars -- do not have jets.

In general, the maximum terminal poloidal velocity is of the 
order of the Keplerian speed at the disk inner radius (see Tab.\,2).
The speed is highest along the field lines from the gap due to the
low flow density.
The axial flow speed is low. 
Its mass density depends on the injection parameters and
could be relatively large (S4, L5) or low (L3, S2), but is
generally less than 10\% of the density at the inner disk 
radius.
The maximum stellar wind speed is reached along field lines with the
largest opening angle and is above the Keplerian speed at the
disk inner radius (Fig.\,5).
The same holds for the maximum disk wind speed. However, in this case
the maximum speed is reached along the field lines with a foot point 
at small radius.
This is partly due to length of the acceleration distance, partly 
due to the rapid rotation of the disk material at small radii.

%---------------------------------------------------------------
\subsubsection{The question of dipolar accretion}
In all our different simulation runs we never observe a signature of
dipolar accretion as it is would be expected from models of young
stellar jet formation (e.g. Camenzind 1990, Shu et al. 1994).
Instead, a magnetically driven wind develops from the stellar surface. 
Note that this strong stellar outflow is
{\em permitted} but 
{\em not prescribed} by the inflow boundary condition along the star,
since the inflow velocity is very low.
We emphasize that even in our simulation L1 for a non rotating star 
(Fendt \& Elstner 1999), 
where no stellar wind can develop, no dipolar accretion occurred.
Therefore, we believe that it is not the boundary condition which 
prevails the matter from falling from the disk to the star.
In fact, dipolar accretion has never been observed in the literature
of numerical MHD simulations considering the star-disk interaction
(e.g. Hayashi et al. 1996, Goodson et al. 1997, Miller \& Stone 1997),
but this might also be caused by the comparatively short time evolution
in those simulations.

We think that the main reason that hinders the dipolar accretion is
the choice of the co-rotation radius equal to the inner disk radius.
Only disk material orbiting inside the co-rotation radius could
be accreted along the field lines.
 
It is clear that such an accretion process along converging
field lines is difficult to treat numerically.
Certainly, this important topic has to be investigated more
deeply. We defer this to a future paper.

%---------------------------------------------------------------
\subsubsection{The question of collimation}
The quasi-stationary two-component outflow obtained in our
simulations show almost {\em no indication for collimation}.
The seems to be in contradiction to the literature 
(OP97, Romanova et al. 1997).
However, the non-collimation of the flows can be explained following the 
analysis of Heyvaerts \& Norman (1989).
They have shown that only jets carrying a net poloidal current will
collimate to a cylindrical shape.
However, in our case we have an initially dipole type magnetosphere
and the final state of a spherically radial outflow enclosing a 
neutral line with a poloidal magnetic field reversal. 
The toroidal field reversal also implies a reversal of the poloidal current
density with only a weak {\em net} poloidal current. 
In such a configuration, a self-collimation of the flow as 
obtained by OP97 or Romanova et al. (1997) cannot be achieved.
In both of these publications a net poloidal current flows along
the {\em monotonously} distributed field lines.

One might expect to obtain the OP97 results of a collimating jet 
as a limiting case in the present simulations, just due to the
fact that the inflow boundary condition along the disk is the
same as in their setup.
However, we think that this is not possible since the initial field 
structure and, thus, the flux distribution in the lower boundary is 
completely different.
A combination of the OP97 field distribution together with a central
dipole might do the job.
Still, the problem would be the very different field strength of both 
components, since the dipolar field will decrease by a factor of 10-50
towards the inner disk radius. 
Thus, the stellar field is always dominating the numerical simulation.
We defer the treatment of such a completely new numerical setup
to a future paper.

Apart from this argument concerning a flow self-collimation we 
mention the hypothesis raised by Spruit et al. (1997) claiming that a
``poloidal collimation'' is responsible for the jet structure.
Such a poloidal collimation would rely on the magnetic pressure onto
the jet flow from the disk magnetic field {\em outside} the jet.
Their condition for poloidal collimation, a disk magnetic field
distribution $\bp \sim r^{-q}$ with $q\leq 1.3$, is clearly not satisfied
in our case of a dipole type field distribution along the disk
(which is conserved from the initial condition because of flux
conservation). 
For a dynamo generated field in the disk this condition is satisfied
(R\"udiger et al. 1995). This holds also for the disk field distribution
of OP97.
In this sense, our simulations are consistent with Spruit et al. (1997),
although we do not argue that our results support their hypothesis
that ``poloidal collimation'' is the main process to produce jets.

%---------------------------------------- Figure 8
\setlength{\unitlength}{1mm}
\begin{figure}
\parbox{100mm}
\thicklines
\epsfxsize=40mm
%\put(0,160){\epsffile{9822f8a.ps}}
%\put(45,160){\epsffile{9822f8b.ps}}
%\put(0,120){\epsffile{9822f8c.ps}}
%\put(45,120){\epsffile{9822f8d.ps}}
%\put(0,80){\epsffile{9822f8e.ps}}
%\put(45,80){\epsffile{9822f8f.ps}}
%\put(0,40){\epsffile{9822f8g.ps}}
%\put(45,40){\epsffile{9822f8h.ps}}
%\epsfxsize=43mm
%\put(-3,-5){\epsffile{9822f8i.ps}}
%\put(42,-5){\epsffile{9822f8j.ps}}
%
\caption
{Properties of the stationary final state for the example
solution L5.
The time step is $t = 200$ when a stationary state has been
reached.
Conserved quantities of stationary MHD:
magnetic flux function $\Psi$ (poloidal field lines),
mass flux per flux surface $\eta(r,z)\equiv \eta(\Psi)$,
the iso-rotation parameter $\Omega_F(r,z)\equiv \Omega_F(\Psi)$, 
the total angular momentum $L(r,z)\equiv L(\Psi)$.
Density distribution $\rho(r,z)$,
poloidal electric current $j_p(r,z)$, 
angular velocity $\Omega (r,z)$,
toroidal magnetic field strength $B_{\phi}$ (positive (negative) values
outside (inside) the neutral layer),
vectors of poloidal Lorentz force $F_{L,p}(r,z)$ and
Lorentz force component parallel to the poloidal field lines
$F_{L,||}(r,z)$.
(Vectors are normalized to unity).
}
\end{figure}

We further to note the results of Ustyugova et al. (1999) who claim
that the shape of the numerical box influences the degree of
collimation. 
A rectangular box extended along the symmetry axis would lead to an
artificial flow collimation, 
whereas a quadratic box simulation (as used in our simulation)
did not result in a collimated structure.
A recent study by Okamoto (1999) also has raised strong arguments
against a MHD self-collimation. 
In particular, he claims that {\em electric current-closure} will
inhibit a self-collimation, 
a point which is not always considered in MHD jet models.
Current-closure, however, is satisfied in our model due to the
reversed dipole type initial field.

Nevertheless, strongly collimated astrophysical jet flows are 
observed.
For the moment we speculate that an increase of the 
{\em disk field strength}
would probably enhance the degree of collimation.
So far we doubt whether an increase of the size of the computational 
box will be sufficient,
because in our model the field distribution and mass flow rate 
decrease strongly with radius.

%---------------------------------------------------------------
\subsubsection{The final steady state - application of stationary MHD}
It is a well known from standard MHD theory that an axisymmetric 
stationary MHD flow is defined by five integrals of motion along the 
magnetic flux function 
$\Psi(r,z) \equiv 2\pi \int \vec{B}_{\rm P}\cdot d\vec{A}$.
Stationarity implies the following conserved quantities along the 
flux surface $\Psi$:
The mass flow rate per flux surface, 
$\eta(\Psi)\equiv {\rm sign}(\vec{v}_{\rm P}\cdot \vec{B}_{\rm P}) 
\,\rho v_{\rm P}/B_{\rm P}$,
the iso-rotation parameter 
$\Omega_{\rm F}(\Psi) \equiv (v_{\phi} - \eta B_{\phi}/\rho)/r $,
the total angular momentum density per flux surface,
$L(\Psi) \equiv r ( v_{\phi} - B_{\phi}/\eta) $,
and the total energy density $E(\Psi)$.
Therefore, for a time-dependent axisymmetric simulation evolving 
into a stationary state,
these functions must be constant along the field lines.
Figure 8 demonstrates such a behavior for the example solution
L5 for the quantities $\eta(\Psi)$,$\Omega_{\rm F}(\Psi)$ and 
$L(\Psi)$.
An overlay of each of these functions 
with the contour plot of the field lines $\Psi(r,z)$ 
would show that the contours are perfectly `parallel'.
This clearly proves the stationary character of the final state of
simulation L5.
For comparison, Fig.\,8 shows the distribution of the poloidal electric
current density, the toroidal magnetic field, and the angular velocity
of the plasma.

For a stationary solution it is interesting to investigate the
Lorentz force projected parallel and perpendicular to the field 
lines.
The poloidal Lorentz force vectors (Fig.\,8) show that in the
region of the highest poloidal velocities, the Lorentz force is
more or less aligned with the field lines.
In the region of low poloidal velocity the main component of
the Lorentz force is perpendicular to the field lines.
As discussed above (Sect.\,4.4.2.) the perpendicular component of the
Lorentz force acts collimating inside the neutral layer and de-collimating
outside the neutral layer.

%---------------------------------------- Figure 9
\setlength{\unitlength}{1mm}
\begin{figure*}
\parbox{100mm}
\thicklines
\epsfysize=5mm
%\put(10,80){\epsffile{9822f9a.ps}}
%
\epsfysize=60mm
%
%\put(10,10){\epsffile{9822f9b.ps}}
%\put(85,10){\epsffile{9822f9c.ps}}
%
\caption
{Density distribution 
and poloidal magnetic field lines 
of simulation L5 for initial time step ({\em left})
and for the final stationary state ({\em right})
plotted for all four hemispheres.
}
\end{figure*}

Another interesting feature is the {\em direction} of the
parallel component of the Lorentz force (see Fig.\,8).
Close to the disk boundary there is a region where this component 
changes sign and the Lorentz force is 
therefore {\em decelerating} the matter\footnote{In Fig.\,8, this 
region is only one vector element wide. However, a better resolution
shows that this region extends to about $z=1$ along the outer disk}.
This demonstrates that in the stationary final state the Lorentz force
is {\em not} the main driving mechanism of the disk material from
the disk into the corona.
It is only at a larger height above the disk that the parallel
component of the Lorentz force accelerates the plasma.

As a summary of this section we show in Fig.\,9 the final
stationary state of the example solution L5 plotted for all
four hemispheres. 
This figure (and only this one) is {\em rotated by $90\degr$} with
the $z$-axis in vertical direction.
This gives a comparative look how the simulation has evolved from the
initial dipolar structure to the spherically radial outflow final 
state.

%---------------------------------------------------------------
\section{Summary}
%---------------------------------------------------------------
%
We have performed numerical simulations of the evolution of a 
stellar dipole type magnetosphere in interaction with a 
Keplerian accretion disk
using the ideal MHD ZEUS-3D code in the axisymmetry option.
The simulations are lasting over hundreds (or even thousands)
of rotational periods of the inner disk.
The central star is rotating with a co-rotation radius chosen
as the disk inner radius.
A smooth mass inflow is prescribed into the corona which is initially
in hydrostatic equilibrium.
The initial dipole type magnetic field distribution is force-free.
The density and velocity profile as well as the magnetic
field profile along the inflow boundary has not been changed during
the computation.

Our main results are summarized as follows.
\begin{itemize}
\item[(1)] The initial dipolar field breaks up by a combined action
  of the winding-up process due to differential rotation between the
  star and disk and the wind mass loss from star and disk.
  `Stellar' and 'disk' field lines remain disconnected after the
  disrupting `bubble' has left the computational grid.
\item[(2)] A two-component MHD wind leaves both the disk and the
  rotating star moving away in radial direction.
  The two components are divided by a neutral field line. 
  The magnetic field direction (both, poloidal and toroidal)
  is positive outside the neutral line and negative inside.
  This field reversal is a characteristic difference from
  jet simulations of OP97 and Romanova et al. 1997.
\item[(3)] Mainly dependent on the wind mass flow rates a stationary or
  quasi-stationary state evolves after hundreds or thousands of inner
  disk rotations.
  The disk wind always evolves into a stationary state.
  A high stellar wind mass loss rate supports `complete stationarity',
  i.e. stationarity also for the stellar wind component.
\item[(4)] The initial driving mechanism of the disk wind are centrifugal
  forces of the rotation matter leaving the disk in vertical direction.
  At larger heights above the disk, this matter becomes magnetically
  accelerated.
  The maximum flow speed is about the Keplerian velocity at the inner disk 
  radius. The hight speed is observed in the outer layers of the stellar 
  wind and in the upper layers of the disk wind.
\item[(5)] Depending on the stellar wind mass flow rate, knots may form
  along the symmetry axis.
  The size of these knots is about the inner disk radius, but also depends 
  weakly on the grid resolution.
  The knot pattern velocity is about 10\% of the Keplerian speed at
  the disk inner radius.
  The full opening angle of `knot flow` is about 30\degr to 50\degr.
  Concerning the knot's size and spacing, it is unlikely that these 
  features are correlated to the observed knots in protostellar jets, 
  but may be connected to QPO's in X-ray binaries.
\item[(6)] There is almost no indication for a flow self-collimation. 
  The flow structure remains more or less conical. 
  We believe that the main reason for the lack of collimation is the field
  reversal between stellar and disk wind also implying a reversal in
  the poloidal current density. 
  Thus, the net poloidal current will be weak.
  This is a major difference to OP97 and Romanova et al. 1997.
  However, this result could also interpreted in terms of a missing 
  poloidal collimation mechanism proposed by Spruit et al. (1997).
\item[(7)] No signature of an accretion stream along a dipolar field
  channel towards stellar surface is observed. This may be due to the 
  fact that the dipolar field has completely disappeared or due to our
  choice of the co-rotation radius.
\end{itemize}

Our results are in general applicable to any star-disk system which
is coupled by magnetic fields.
One critical aspect may be that we assume a fixed boundary condition
for the magnetic field in the disk.
However, if the field structure in the corona changes as drastically
as we have shown, this might influence also the magnetic flux
distribution in the disk.
But then, for a proper time-dependent disk boundary condition, the disk 
structure should be treated in a more detailed manner. 
This however, is beyond the scope of the present paper.
From our results we like to put forward the following main hypotheses.
\begin{itemize}
\item[(8)] Star-disk systems are supposed to have a two component
  wind/jet structure.
\item[(9)] A strong stellar field (equivalent to a low stellar mass
  loss rate)
  leads to instabilities along the rotational axis.
  A strong and stable jet is unlikely in such objects. 
  This may be one reason why highly magnetized stars with disks like
  neutron stars or magnetic white dwarfs do {\em not} have jets. 
  A disk field generated by a turbulent dynamo could be a better 
  candidate for driving the jet.   
\item[(10)] The current model of magnetized accretion in young stars
  along dipolar field lines from disk to star have to be re-considered.
  The magnetospheric structure often inferred from stationary models
  (Camenzind 1990; Shu et al. 1994)
  may completely change if the time-dependent evolution is considered.
\end{itemize}

%---------------------------------------------------------------
\begin{appendix}
%---------------------------------------------------------------
%
%---------------------------------------------------------------
\section{Numerical tests}
%---------------------------------------------------------------
%
Here, we will discuss numerical test solutions.
The first test example is the recalculation of the OP97 solution of
an axisymmetric jet propagating from a rotating Keplerian disk.
Such a  scenario is similar to the one treated in the present paper,
however, with a parabolic-type initial potential field configuration
and without a central rotating star.
Figure A.1 shows the result of our simulation at $t=100$.
At this time the jet bow shock has traveled 52 units along the
z-axis.
In the long-term evolution, the location of the magneto-sonic surfaces
agrees with the results of OP97.
However, as a (minor) difference we note that
for our test simulations fitting to the OP97 results
the value for the plasma-beta is smaller by about a factor of
$\sqrt{4\pi}$ compared to OP97. 
We use the original ZEUS code normalized with $P_{\rm gas} \sim B^2/2$
(instead of the usual $P_{\rm gas} \sim B^2/8\pi$) (user manual).
Therefore, in order to match the definition of the plasma beta as
$\beta_i = 8\pi P_i /B^2_i$, one must define the field strength
properly (this gives a factor of $\sqrt{4\pi}$ in the field strength).
Thus, the difference in the $\beta_i $ may be due to a different 
normalization applied in OP97 (Ouyed 2000, private communication).
We have deduced this factor by {\em comparing both simulations}.
Additionally, it becomes understandable with the normalization of
the ZEUS code magnetic field.
For $\beta_i = 1.0$ in {\em our} simulations the jet solution is appropriate
slower, reaching only $z=42$ after 100 time steps
\footnote{This is, by the way, remarkably similar to the location of the
shock front in the velocity plot (Fig.\,6) in OP97, which is 
different from the one in their density plot (Fig.\,3).
However, we note that a `wrong' plasma-beta must be visible in the
location of the magneto-sonic surfaces.  This is not the case.
Therefore, we conclude the differences in $\beta_i$ are just due to 
a different field normalization in the actual codes}.
Concluding that our recalculation of the OP97 model was successful, we
note however the tiny `wave' pattern of our density contours (Fig.\,A.1).
This wave pattern is present only in the hydrodynamic variables,
but not in the magnetic field.
These density variations are less then 10\%.

%---------------------------------------- Figure A1
\setlength{\unitlength}{1mm}
\begin{figure*}
\parbox{100mm}
\thicklines
\epsfysize=60mm
%\put(0,0){\epsffile{9822fa1a.ps}}
\epsfysize=60mm
%\put(60,0){\epsffile{9822fa1b.ps}}
\epsfysize=50mm
%\put(120,5){\epsffile{9822fa1c.ps}}
%
\caption
{Numerical test example.
Re-calculation of the OP97 jet model.
From {\it left} to {\it right}:
Poloidal magnetic field lines, density contours, 
and poloidal velocity vectors 
at $t= 100$, calculated with $\beta_i = 0.28$.
The location of the shock front is the same for all three plots
(in difference to OP97). 
}
\end{figure*}

As a second example for a numerical test, we show an overlay
of the density contours and poloidal field lines of six initial
time steps of the simulation L1, {\em before} the torsional Alfv\'en
wave has passed the outer (upper) grid boundary.
It can be seen that upstream of the bow shock front the
magneto-hydrostatic initial condition remains in perfect equilibrium.
Thus, force-freeness of the initial magnetic field as well as
the hydrostatic equilibrium is satisfied with good accuracy.
Without the inflow boundary condition at $z=0$ the initial equilibrium
will remain unchanged.

%---------------------------------------- Figure A2
\setlength{\unitlength}{1mm}
\begin{figure}
\parbox{100mm}
\thicklines
\epsfysize=60mm
%\put(0,0){\epsffile{9822fa2a.ps}}
%\put(0,60){\epsffile{9822fa2b.ps}}
%
\caption
{Numerical test example. Solution L1 without stellar rotation.
Overlay of a couple of initial time steps
(at $t= 0, 10, 20, 30, 40, 50 $).
Poloidal magnetic field lines ({\it top})
and density contours ({\it bottom}).
Thick lines indicate the initial distributions.
The solutions perfectly match in the regions not yet 
disturbed by the inflow boundary condition.
The long term evolution of this solution is shown in
FE99.
}
\end{figure}

%---------------------------------------------------------------
\end{appendix}
%---------------------------------------------------------------
 
\begin{acknowledgements}
Encouraging discussions with R.~Pudritz and R.~Ouyed are acknowledged
by C.F.
We thank the LCA team and M.~Norman for the possibility to use the
ZEUS code.
\end{acknowledgements}

%---------------------------------------------------------------

%---------------------------------------------------------------
%---------------------------------------------------------------
\end{document}